\documentclass[aps,pre,showpacs,twocolumn]{revtex4-1}

\usepackage[T1]{fontenc}
\usepackage{amssymb}
\usepackage{amsbsy}
\usepackage{amsmath}
\usepackage{graphicx}
\usepackage{hyperref} 

\begin{document}


\def\beq{\begin{equation}}
\def\eeq{\end{equation}}
\def\llbrace{\left\lbrace}
\def\rrbrace{\right\rbrace}
\def\lbraket{\left[}
\def\rbraket{\right]}

\newcommand{\Tr}{{\rm Tr}} 
\newcommand{\tr}{{\rm tr}} 
\newcommand{\sgn}{{\rm sgn}} 
\newcommand{\mean}[1]{\langle #1 \rangle}
\newcommand{\const}{{\rm const}} 

\newcommand{\cc}{{\rm c.c.}} 
\newcommand{\hc}{{\rm h.c.}} 

\def\eps{\epsilon}
\def\gam{\gamma} 
\def\phibf{\boldsymbol{\phi}}
\def\varphibf{\boldsymbol{\varphi}}
\def\psibf{\boldsymbol{\psi}}
\def\lamb{\lambda}
\def\sig{\sigma}
\def\sigp{{\sigma'}} 
\def\bfsig{\boldsymbol{\sigma}} 
\def\sigbf{\boldsymbol{\sigma}}

\def\half{\frac{1}{2}}
\def\dhalf{\dfrac{1}{2}}
\def\third{\frac{1}{3}} 
\def\quarter{\frac{1}{4}}

\def\h{{\bf h}}
\def\n{{\bf n}} 
\def\p{{\bf p}} 
\def\q{{\bf q}}
\def\r{{\bf r}}
\def\v{{\bf v}}
\def\A{{\bf A}}

\def\P{{\bf P}} 
\def\Q{{\bf Q}} 
\def\nablabf{\boldsymbol{\nabla}}
\def\Pibf{\boldsymbol{\Pi}}
\def\pibf{\boldsymbol{\pi}}

\def\para{\parallel}

\def\w{\omega}
\def\wn{\omega_n}
\def\wnu{\omega_\nu}
\def\wp{\omega_p} 
\def\dmu{{\partial_\mu}}
\def\dl{{\partial_l}}  
\def\dt{\partial_t} 
\def\dk{\partial_k}
\def\dtau{{\partial_\tau}}  
\def\det{{\rm det}} 

\def\intr{\int d^dr}  
\def\inttau{\int_0^\beta d\tau}
\def\inttaup{\int_0^\beta d\tau'}
\def\inttaur{\int_0^\beta d\tau \int d^dr}
\def\intinf{\int_{-\infty}^\infty}
\def\intw{\int_{-\infty}^\infty \frac{d\w}{2\pi}}
\def\sumr{\sum_{\bf r}} 

\def\calC{{\cal C}} 
\def\dt{\partial_t}
\def\calD{{\cal D}}
\def\calO{{\cal O}}

\def\Sign{\Sigma_{\rm n}} 
\def\Sigan{\Sigma_{\rm an}} 
\def\Signt{\tilde\Sigma_{\rm n}} 
\def\Sigant{\tilde\Sigma_{\rm an}} 
\def\Gn{G_{\rm n}}
\def\Gan{G_{\rm an}}
\def\Itildell{\tilde I_{k,ll}}
\def\Itildett{\tilde I_{k,tt}}
\def\Jtildellll{\tilde J_{k;ll,ll}}
\def\Jtildetttt{\tilde J_{k;tt,tt}}
\def\Jtildeltlt{\tilde J_{k;lt,lt}}
\def\Jtildelltt{\tilde J_{k;ll,tt}}
\def\Jtildettll{\tilde J_{k;tt,ll}}
\def\Jtildelllt{\tilde J_{k;ll,lt}}
\def\Jtildeltll{\tilde J_{k;lt,ll}}
\def\Jtildettlt{\tilde J_{k;tt,lt}}
\def\Jtildelttt{\tilde J_{k;lt,tt}}


\title{Infrared behavior in systems with a broken continuous symmetry: classical O($N$) model vs interacting bosons}

\author{N. Dupuis}
\affiliation{Laboratoire de Physique Th\'eorique de la Mati\`ere Condens\'ee, 
CNRS - UMR 7600, \\ Universit\'e Pierre et Marie Curie, 4 Place Jussieu, 
75252 Paris Cedex 05,  France }

\date{January 14, 2010}

\begin{abstract}
In systems with a spontaneously broken continuous symmetry, the perturbative loop expansion is plagued with infrared divergences due to the coupling between transverse and longitudinal fluctuations. As a result the longitudinal susceptibility diverges and the self-energy becomes singular at low energy. We study the crossover from the high-energy Gaussian regime, where perturbation theory remains valid, to the low-energy Goldstone regime characterized by a diverging longitudinal susceptibility. We consider both the classical linear O($N$) model and interacting bosons at zero temperature, using a variety of techniques: perturbation theory, hydrodynamic approach (i.e., for bosons, Popov's theory), large-$N$ limit and non-perturbative renormalization group. We emphasize the essential role of the Ginzburg momentum scale $p_G$ below which the perturbative approach breaks down. Even though the action of (non-relativistic) bosons includes a first-order time derivative term, we find remarkable similarities in the weak-coupling limit between the classical O($N$) model and interacting bosons at zero temperature.  
\end{abstract}
\pacs{05.30.Jp,05.70.Fh}

\maketitle

\section{Introduction}

In the context of critical phenomena, it is well known that the Gaussian approximation breaks down in the vicinity of a second-order phase transition (below the upper critical dimension). When the Ginzburg criterion $|T-T_c|/T_c\gg t_G$ is violated ($T_c$ denotes the critical temperature and $|T-T_c|/T_c\sim t_G$ defines the Ginzburg temperature $T_G$), the long-distance behavior of the correlation functions cannot be described by a Gaussian fluctuation theory and more involved techniques, such as the renormalization group, are required~(see e.g.~\cite{Ma_book}). At the critical point ($T=T_c$), one can nevertheless distinguish two regimes in momentum space: a high-energy Gaussian regime, where the Gaussian approximation remains essentially correct, and a low-energy critical regime where the correlation function of the order parameter field shows a critical behavior characterized by a non-zero anomalous dimension $\eta$. These two regimes are separated by a characteristic momentum scale $p_G$ which defines the Ginzburg length $\xi_G=p_G^{-1}$~(see e.g.~\cite{Chaikin_book}).  

In systems with a broken continuous symmetry, the physics remains non-trivial in the whole low-temperature phase due to the presence of Goldstone modes, which implies that correlations decay algebraically. The coupling between transverse and longitudinal order parameter fluctuations leads to a divergence of the longitudinal susceptibility~\cite{Patasinskij73,Fisher73,Anishetty99}. Away from the critical regime (i.e. at sufficiently low temperatures), one can distinguish a high-energy Gaussian regime ($|\p|\gg p_G$), where the Gaussian approximation remains correct, and a low-energy Goldstone regime ($|\p|\ll p_G$) dominated by the Goldstone modes and characterized by a divergence of the longitudinal susceptibility. Note that the Ginzburg momentum scale $p_G$ defined here is the same as the one signaling the onset of the critical regime (in momentum space) when the system is near the phase transition. For instance, for the $(\varphibf^2)^2$ theory with O($N$) symmetry (classical O($N$) model), one finds a transverse susceptibility $\chi_\perp(\p)\sim 1/\p^2$ for $\p\to 0$, while the longitudinal susceptibility $\chi_\para(\p)\sim 1/|\p|^{4-d}$ is also singular in dimensions $2<d\leq 4$ (the divergence is logarithmic for $d=4$). At and below the lower critical dimension $d_c^-=2$, transverse fluctuations lead to a suppression of long-range order (Mermin-Wagner theorem). There is an analog phenomenon in zero-temperature quantum systems with a broken continuous symmetry. When the Goldstone mode frequency $\w=c|\p|$ vanishes linearly with momentum, the longitudinal susceptibility $\chi_\para(\p,\w)\sim 1/(\w^2-c^2\p^2)^{(3-d)/2}$ has no pole-like structure but a branch-cut for $d\leq 3$, and the dynamical structure factor exhibits a critical continuum above the usual delta peak $\delta(\w-c|\p|)$ due to the Goldstone mode~\cite{Sachdev99,Zwerger04,Dupuis09b}. 

Historically, the divergence of the longitudinal susceptibility was encountered (although not recognized as such) early on in interacting boson systems. The first attempts to improve the Bogoliubov theory of superfluidity~\cite{Bogoliubov47} were made difficult by a singular perturbation theory plagued by infrared divergences~\cite{Beliaev58a,Beliaev58b,Hugenholtz59,Gavoret64}. As realized later on~\cite{Nepomnyashchii75,Nepomnyashchii78,Nepomnyashchii83}, the singular perturbation theory is a direct consequence of the coupling between transverse and longitudinal fluctuations.

In this paper, we study the crossover from the high-energy Gaussian regime to the low-energy Goldstone regime in the ordered phase, both for the classical O($N$) model and interacting bosons at zero temperature. Even though the action of (non-relativistic) bosons includes a first-order time derivative term, which prevents a straightforward description in terms of a classical O(2) model, we find remarkable similarities in the weak-coupling limit between these two models. On the other hand, the strong-coupling limit of the O($N$) model, i.e. the critical regime near the phase transition, has no direct analog in zero-temperature interacting boson systems. 

The classical O($N$) model is studied in Sec.~\ref{sec_phi4}, while superfluid systems are discussed in Sec.~\ref{sec_bosons}. First, we show that the loop expansion about the mean-field solution is plagued with infrared divergences and deduce a perturbative estimate of the Ginzburg momentum scale $p_G$ (Secs.~\ref{subsec_phi4_pt} and \ref{subsec_bosons_pt}). Then, we use symmetry arguments to derive the exact value of the self-energies at vanishing momentum (and frequency) (Secs.~\ref{subsubsec_phi4_sigmaexact} and \ref{subsubsec_bosons_sigmaexact}). In the case of bosons, we obtain Nepomnyashchii and Nepomnyashchii's result about the vanishing of the anomalous self-energy~\cite{Nepomnyashchii75}. In Secs.~\ref{subsec_phi4_hydro} and \ref{subsec_bosons_hydro}, we show that the difficulties of perturbation theory can be circumvented within a hydrodynamic approach (i.e., for bosons, Popov's theory~\cite{Popov72,Popov_book_2,Popov79}) based on an amplitude-direction representation of the order parameter field. This yields the correlation functions in the hydrodynamic regime defined by a characteristic momentum scale $p_c\gg p_G$. The O($N$) model is solved in the large-$N$ limit in Sec.~\ref{subsec_largeN}. This allows us to obtain the longitudinal correlation function in the whole low-temperature phase, including the critical regime in the vicinity of the phase transition. Finally, we show how the non-perturbative renormalization group (NPRG) provides a natural framework to understand the ordered phase of the O($N$) model and the superfluid phase of interacting bosons (Secs.~\ref{subsec_phi4_nprg} and \ref{subsec_bosons_nprg}).

\section{The $(\varphibf^2)^2$ theory at low temperatures} 
\label{sec_phi4}

We consider the $(\varphibf^2)^2$ theory defined by the action
\beq
S[\varphibf] = \intr \llbrace \half (\nablabf\varphibf)^2 + \frac{r_0}{2} \varphibf^2 + \frac{u_0}{4!} (\varphibf^2)^2 \rrbrace  
\label{action} 
\eeq
where $\varphibf$ is a $N$-component real field and $d$ the space dimension. We assume $N\geq 2$ and $d>2$. The model is regularized by a ultraviolet momentum cutoff $\Lambda$. The connected propagator 
\beq
G_{ij}(\p) = \mean{\varphi_i(\p) \varphi_j(-\p)} - \mean{\varphi_i(\p)} \mean{\varphi_j(-\p)}
\eeq
is related to the self-energy $\Sigma$ by Dyson's equation $G^{-1}=G_0^{-1}+\Sigma$, where 
\beq
G_{0,ij}(\p) = \frac{\delta_{i,j}}{\p^2+r_0} 
\eeq
is the bare propagator. In the low-temperature phase, if we denote by $\varphibf_0=\mean{\varphibf(\r)}$ the order parameter, the self-energy
\begin{align}
\Sigma_{ij}(\p) &= \hat\varphibf_{0,i} \hat\varphibf_{0,j} \Sigma_l(\p) + (\delta_{i,j}-\hat\varphibf_{0,i} \hat\varphibf_{0,j}) \Sigma_t(\p) \nonumber \\ 
&= \delta_{i,j} [\Sign(\p) - \Sigan(\p)] + 2 \hat\varphibf_{0,i} \hat\varphibf_{0,j} \Sigan(\p) 
\label{sigma1}
\end{align} 
($\hat\varphibf_0=\varphibf_0/|\varphibf_0|$) can be written in terms of its longitudinal ($\Sigma_l$) and transverse ($\Sigma_t$) parts. In the second line of (\ref{sigma1}), we have introduced the ``normal'' ($\Sign$) and ``anomalous''  ($\Sigan$) self-energies. In the following, we assume that the order parameter $\varphibf_0$ is along the direction $(1,0,\cdots,0)$ so that 
\beq
\Sigma_{ii}(\p) = \llbrace  
\begin{array}{ccc}
\Sign(\p) + \Sigan(\p) & \mbox{if} & i=1 , \\  
\Sign(\p) - \Sigan(\p) & \mbox{if} & i \neq 1 . 
\end{array}
\right. 
\eeq

The anomalous self-energy $\Sigan$ is related to the spontaneously broken O($N$) symmetry and vanishes in the high-temperature phase. $\Sign$ and $\Sigan$ are analogous to the normal and anomalous self-energies which are usually introduced in the theory of superfluidity~\cite{AGD_book,Fetter_book}. For $N=2$, we can introduce the complex field
\beq
\psi(\r) = \frac{1}{\sqrt{2}}[\varphi_1(\r) + i\varphi_2(\r)] .
\eeq
Making use of the two-component field 
\beq
\Psi(\r) = \left( \begin{array}{c} \psi(\r) \\ \psi^*(\r) \end{array} \right) , \quad 
\Psi^\dagger(\r) = \left( \psi^*(\r), \psi(\r) \right) ,
\eeq
the two-point propagator becomes a $2\times 2$ matrix in Fourier space, whose inverse is given by 
\beq
\left( \begin{array}{cc} \p^2+r_0+\Sign(\p) & \Sigan(\p) \\ \Sigan(\p) & \p^2+r_0+\Sign(\p) \end{array} \right) ,
\eeq
and bears some similarities with the single-particle propagator in a superfluid (Sec.~\ref{sec_bosons}).

\subsection{Gaussian approximation and breakdown of perturbation theory}
\label{subsec_phi4_pt}

Let us begin with a dimensional analysis of the action (\ref{action}). If we assign the scaling dimension 1 to momenta (i.e. $[\p]=1$), the field has engineering dimension $[\varphibf]=\frac{d-2}{2}$, $[r_0]=2$ and $[u_0]=4-d$. We can then define two characteristic length scales,
\beq
\begin{split}
\xi &\sim |r_0|^{-1/2} , \\ 
\xi_G &\sim u_0^{1/(d-4)} .
\end{split}
\label{diman}
\eeq
In the critical regime of the low-temperature phase ($\xi\gg\xi_G$), $\xi_G$ is the characteristic length scale associated to the onset of critical fluctuations, while $\xi\equiv\xi_J$ is the Josephson length separating the critical regime from a regime dominated by Goldstone modes~\cite{Josephson66}. When critical fluctuations are taken into account, one finds that $\xi_J$ diverges with a critical exponent $\nu$ which differs from the mean-field value $1/2$. At low temperatures away from the critical regime ($\xi\ll\xi_G$), $\xi\equiv\xi_c$ corresponds to a correlation length for the gapped amplitude fluctuations while direction fluctuations are gapless due to Goldstone's theorem. The physical meaning of the Ginzburg length $\xi_G$ in this temperature range will become clear below.

\subsubsection{Gaussian approximation}

Within the mean-field (or saddle-point) approximation, one finds $\varphi_0=|\varphibf_0|=(-6r_0/u_0)^{1/2}$ in the low-temperature phase ($r_0<0$). In the Gaussian approximation, one expands the action to quadratic order in the fluctuations $\varphibf-\varphibf_0$~\cite{Ma_book}. This yields the (zero-loop) self-energy
\beq
\Sigma^{(0)}_{ii}(\p) = \llbrace 
\begin{array}{ccc}
-3r_0 & \mbox{if} & i=1 , \\ 
-r_0 & \mbox{if} & i\neq 1 , 
\end{array}
\right. 
\eeq
from which we obtain the longitudinal and transverse propagators, 
\beq
\begin{split}
G_l^{(0)}(\p) &= G_{11}^{(0)}(\p) = \frac{1}{\p^2+2|r_0|} , \\
G_t^{(0)}(\p) &= G_{22}^{(0)}(\p) = \frac{1}{\p^2} .  
\end{split}
\eeq
In agreement with Goldstone's theorem, the transverse propagator is gapless, whereas the longitudinal susceptibility $G_l(\p=0)=1/|2r_0|$ is finite. We shall see below that this last property is an artifact of the Gaussian approximation.  

\subsubsection{One-loop correction and the Ginzburg momentum scale}

\begin{figure}
\centerline{\includegraphics[height=1.8cm]{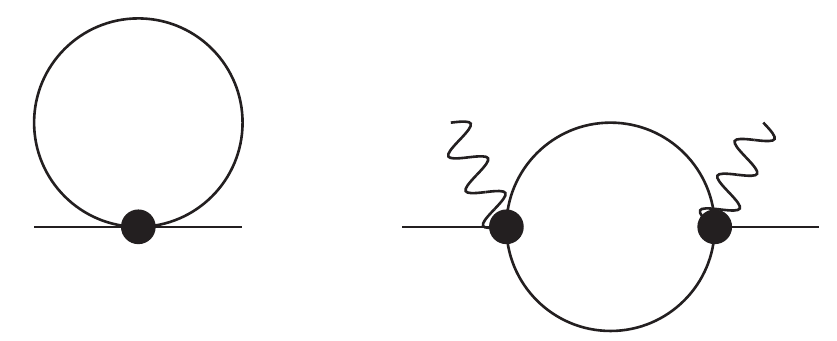}}
\caption{One-loop correction $\Sigma^{(1)}$ to the self-energy. The dots represent the bare interaction, the zigzag lines the order parameter $\varphi_0$, and the solid lines the connected propagator $G^{(0)}$. }
\label{fig_self_1loop} 
\end{figure}

The one-loop correction $\Sigma^{(1)}$ to the self-energy is shown diagrammatically in Fig.~\ref{fig_self_1loop}. While the first diagram is finite, the second one gives a diverging contribution to $\Sigma_{11}$ in the infrared limit $\p\to 0$ when $d\leq 4$. The divergence arises when both internal lines correspond to transverse fluctuations, which is possible only for $\Sigma_{11}$. Thus $\Sigma_{22}$ is finite at the one-loop level and the normal and anomalous self-energies exhibit the same divergence,
\beq
\Sign^{(1)}(\p) \simeq \Sigan^{(1)}(\p) \simeq - \frac{N-1}{36} u_0^2 \varphi_0^2 \int_\q \frac{1}{\q^2(\p+\q)^2} ,
\label{sigma_1loop}
\eeq
where we use the notation $\int_\q=\int\frac{d^dq}{(2\pi)^d}$. The momentum integration in (\ref{sigma_1loop}) gives~\cite{Zinn_book_2}  
\beq
\int_\q \frac{1}{\q^2(\p+\q)^2} = \llbrace 
\begin{array}{ccc}
A_d |\p|^{d-4} & \mbox{if} & d<4 , \\
A_4 \ln(\Lambda/|\p|) & \mbox{if} & d=4 ,
\end{array}
\right. 
\label{mint}
\eeq
for $|\p|\ll\Lambda$, where 
\beq
A_d = \llbrace 
\begin{array}{lcc}
- \frac{2^{1-d}\pi^{1-d/2}}{\sin(\pi d/2)} \frac{\Gamma(d/2)}{\Gamma(d-1)} & \mbox{if} & d<4 ,\\ 
\frac{1}{8\pi^2} & \mbox{if} & d=4 .
\end{array}
\right. 
\eeq
The one-loop correction (\ref{sigma_1loop}) diverges for $\p\to 0$ and the perturbation expansion about the Gaussian approximation breaks down. By comparing the one-loop correction to the zero-loop result, i.e. $|\Sign^{(1)}(\p)|\sim \Sign^{(0)}(\p)$ or $|\Sigan^{(1)}(\p)|\sim \Sigan^{(0)}(\p)$, one can nevertheless extract a characteristic (Ginzburg) momentum scale, 
\beq
p_G \sim \llbrace 
\begin{array}{lcc}
[A_d(N-1)u_0]^{1/(4-d)} & \mbox{if} & d<4 , \\ 
\Lambda \exp\left(\frac{-1}{A_4(N-1)u_0}\right) & \mbox{if} & d=4 ,
\end{array}
\right. 
\label{pG1} 
\eeq
which was obtained previously from dimensional analysis [Eq.~(\ref{diman})]. While the Gaussian or perturbative approach remains valid for $|\p|\gg p_G$, the limit $|\p|\ll p_G$ cannot be studied perturbatively. We shall see in Sec.~\ref{subsec_phi4_hydro} that the breakdown of perturbation theory is due to the coupling between transverse and longitudinal fluctuations.

\subsubsection{Exact results for $\Sign(\p=0)$ and $\Sigan(\p=0)$}
\label{subsubsec_phi4_sigmaexact}  

Although the one-loop correction $\Sigma^{(1)}(\p)$ diverges when $\p\to 0$ for $d\leq 4$, it is nevertheless possible to obtain the exact value of $\Sigma(\p=0)$ using the O($N$) symmetry of the model.

Let us consider the effective action 
\beq
\Gamma[\phibf] = -\ln Z[\h] + \intr\, \h \cdot \phibf
\eeq
defined as the Legendre transform of the free energy $-\ln Z[\h]$ where $\h$ is an external field which couples linearly to the $\varphibf$ field and
\beq
\phi_i(\r) = \frac{\delta\ln Z[\h]}{\delta h_i(\r)} = \mean{\varphi_i(\r)}_{\h} . 
\eeq
The notation $\mean{\cdots}_{\h}$ means that the average value is computed in the presence of the external field $\h$. $\Gamma[\phibf]$ satisfies the equation of state
\beq
\frac{\delta\Gamma[\phibf]}{\delta\phi_i(\r)} = h_i(\r) . 
\eeq
At equilibrium and in the absence of external field, the order parameter $\varphibf_0=\mean{\varphibf(\r)}$ is obtained from the stationary condition of the effective action,
\beq
\frac{\delta\Gamma[\phibf]}{\delta\phi_i(\r)}\biggl|_{\phibf(\r)=\varphibf_0} = 0 . 
\eeq
$\Gamma[\phibf]$ is the generating functional of the one-particle irreducible vertices
\beq
\Gamma^{(n)}_{i_1\cdots i_n}(\r_1,\cdots,\r_n) = \frac{\delta^{(n)}\Gamma[\phibf]}{\delta\phi_{i_1}(\r_1) \cdots \delta\phi_{i_n}(\r_n)}\biggl|_{\phibf(\r)=\varphibf_0}  . 
\eeq
The later fully determine the correlation functions. In particular, the two-point vertex $\Gamma^{(2)}$ is related to the propagator by $\Gamma^{(2)} = G^{-1} = G_0^{-1} + \Sigma$. 

The O($N$) invariance of the action (\ref{action}) implies that the effective action $\Gamma[\phibf]$ is invariant under a rotation of the field $\phibf$. Let us consider the case $N=3$ for simplicity (the following results are easily extended to arbitrary $N$). For an infinitesimal rotation $\phibf\to\phibf+\theta \n\times \phibf$ about the axis $\n$ ($\n^2=1$ and $\theta\to 0$), the invariance of the effective action yields 
\beq
\intr \sum_{ijk} \frac{\delta\Gamma[\phibf]}{\delta \phi_i(\r)} \eps_{ijk} n_j \phi_k(\r) =0 , 
\label{wi}
\eeq
where $\eps_{ijk}$ is the totally antisymmetric tensor. Taking the first-order functional derivative $\delta/\delta\phi_l(\r')$ and setting $\phi_i(\r)=\delta_{i,1}\varphi_0$, we obtain
\beq
\intr \sum_{i,j} \Gamma^{(2)}_{il}(\r,\r') \eps_{ij1} n_j = 0 . 
\eeq
With $\n=(0,0,1)$, this gives
\beq
\Gamma^{(2)}_{22}(\p=0) = r_0 + \Sigma_{22}(\p=0) = 0 ,
\label{goldstone} 
\eeq
where $\Gamma^{(2)}(\p)$ denotes $\Gamma^{(2)}(\p,-\p)$. Equation~(\ref{goldstone}) is a direct consequence of Goldstone's theorem. If we  now take the second-order functional derivative $\delta^{(2)}/\delta\phi_l(\r')\delta\phi_m(\r'')$ of (\ref{wi}) and set $\phi_i(\r)=\delta_{i,1}\varphi_0$, we obtain the Ward identity
\begin{multline}
\sum_{i,j} \left[ \Gamma^{(2)}_{im}(\r',\r'')  \eps_{ijl} + \Gamma^{(2)}_{il}(\r'',\r')  \eps_{ijm}   \right] n_j \\ + \intr \sum_{i,j} \Gamma^{(3)}_{ilm}(\r,\r',\r'') \eps_{ij1} n_j \varphi_0 = 0 . 
\end{multline} 
Choosing $l=2$, $m=1$ and $j=3$, this gives 
\beq
\Gamma^{(2)}_{11}(\r',\r'') - \Gamma^{(2)}_{22}(\r'',\r') - \varphi_0 \intr \Gamma^{(3)}_{221}(\r,\r',\r'') = 0. 
\eeq
Integrating over $\r'$ and $\r''$ and using (\ref{goldstone}), we deduce (in Fourier space) 
\beq
\Gamma^{(3)}_{122}(0,0,0) = \frac{\Gamma^{(2)}_{11}(0,0)}{\sqrt{V} \varphi_0} ,
\label{wi1}                                                                            
\eeq
where $V$ is the volume of the system. 

\begin{figure}
\centerline{\includegraphics[width=8cm]{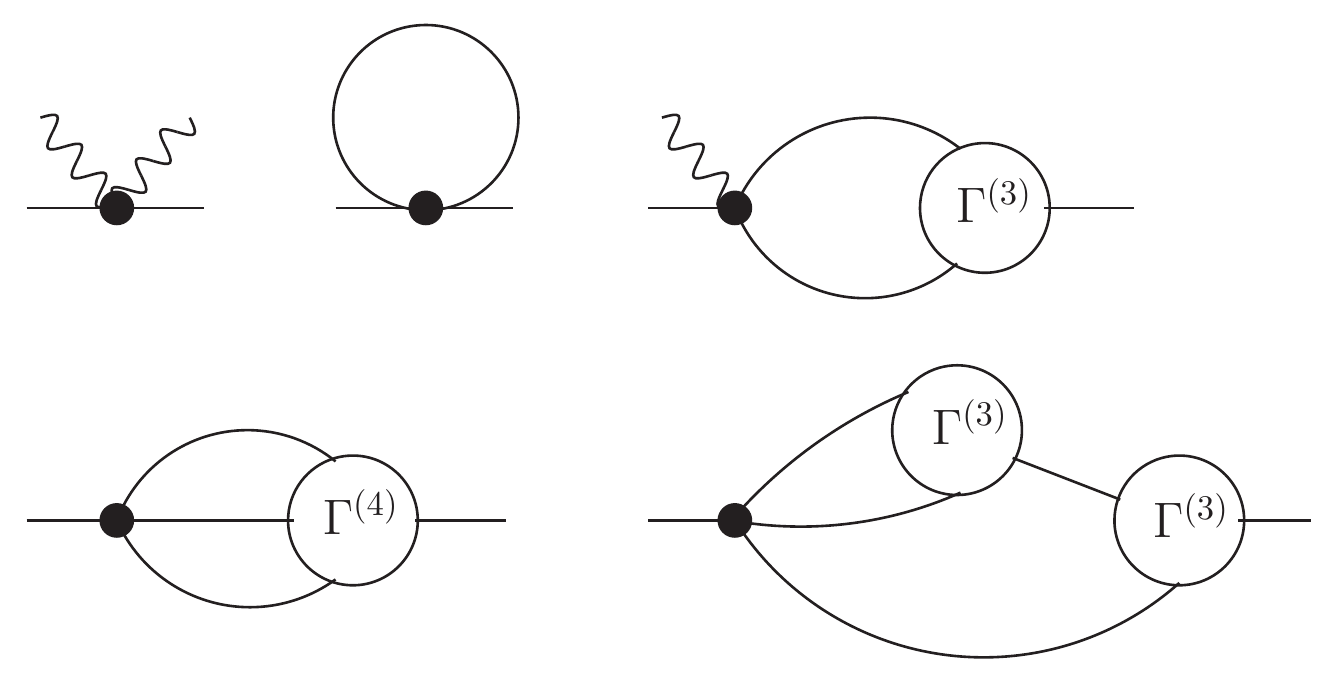}}
\caption{Exact diagrammatic representation of the self-energy in terms of the three- and four-leg vertices $\Gamma^{(3)}$ and $\Gamma^{(4)}$. The dots represent the bare interaction, the zigzag lines the order parameter, and the solid lines the (exact) connected propagator.}
\label{fig_self_exact}
\end{figure}

 Let us now consider the exact diagrammatic representation of the self-energy shown in Fig.~\ref{fig_self_exact}. We know from perturbation theory that the third diagram in Fig.~\ref{fig_self_exact} is potentially dangerous when the two internal lines correspond to transverse fluctuations. We therefore write the self-energy $\Sigma_{11}(\p)$ as 
\begin{align}
\Sigma_{11}(\p) ={}& \tilde\Sigma_{11}(\p) - \frac{N-1}{6\sqrt{V}} u_0\varphi_0 \sum_\q G_{22}(\q) G_{22}(\p+\q) \nonumber \\ & \times \Gamma^{(3)}_{122}(-\p,-\q,\p+\q) ,  
\label{Sig11}
\end{align} 
where $\tilde\Sigma_{11}(\p)$ denotes the part of the self-energy which is regular in perturbation theory (i.e. the part that does not contain pairs of lines corresponding to $G_{22}G_{22}$). If we assume that the transverse propagator $G_{22}(\q)$ is proportional to $1/\q^2$ for $\q\to 0$ (this result will be shown in the following sections), the integral $\int_\q G_{22}(\q)^2$ is infrared divergent for $d\leq 4$. To obtain a finite self-energy $\Sigma_{11}(\p=0)$, one must require that 
\beq
\lim_{\q\to 0}  \Gamma^{(3)}_{122}(0,-\q,\q) = \Gamma^{(3)}_{122}(0,0,0) = 0.
\eeq
The Ward identity (\ref{wi1}) then implies $\Gamma^{(2)}_{11}(\p=0)=0$, so that we finally obtain
\beq
\begin{split}
\Sign(\p=0) &= -r_0 + \half \left[ \Gamma_{11}(\p=0) + \Gamma_{22}(\p=0) \right] = -r_0 , \\
\Sigan(\p=0) &= \half \left[ \Gamma_{11}(\p=0) - \Gamma_{22}(\p=0) \right] = 0 . 
\end{split}
\eeq 
It may appear surprising that the anomalous self-energy, which is related to the spontaneously broken O($N$) symmetry, vanishes for $\p=0$. The equivalent property in interacting boson systems is a fundamental result of the theory of superfluidity (Sec.~\ref{sec_bosons}). 

\subsection{Amplitude-direction representation} 
\label{subsec_phi4_hydro} 

The difficulties of the perturbation theory of Sec.~\ref{subsec_phi4_pt} can be avoided if one uses  the ``good'' hydrodynamic variables in the low-temperature phase, namely the amplitude and the direction of the $\varphibf$ field. We thus write 
\beq
\varphibf(\r) = \rho(\r) \n(\r) , 
\eeq
where $\n(\r)^2=1$, and obtain the action
\beq
S[\rho,\n] = \intr \llbrace \half (\nablabf\rho)^2 + \frac{\rho^2}{2}(\nablabf\n)^2 + \frac{r_0}{2} \rho^2 + \frac{u_0}{4!} \rho^4 \rrbrace .
\eeq
At the mean-field level, the amplitude takes the value $\rho_0=(-6r_0/u_0)^{1/2}$ in the low-temperature phase ($r_0<0$). For small amplitude fluctuations $\rho'=\rho-\rho_0$ (which is expected to be the case at sufficiently low temperatures), we obtain the action 
\beq
S[\rho',\n] = \intr \llbrace \half (\nablabf\rho')^2 + |r_0| \rho'{}^2 + \frac{\rho_0^2}{2} (\nablabf\n)^2  \rrbrace 
\label{action1} 
\eeq
and deduce that the amplitude fluctuations are gapped,
\beq
\mean{\rho'(\p)\rho'(-\p)} = \frac{1}{\p^2+p_c^2} .
\eeq 
If we are interested only in momenta $|\p|\ll p_c=\sqrt{2|r_0|}$, to first approximation we can ignore the higher-order terms in $\rho'$ that were neglected in (\ref{action1}), since they would only lead to a finite renormalization of the coefficients of the action $S[\rho',\n]$~\cite{Zinn_book_2}. 

Equation~(\ref{action1}) shows that in the ``hydrodynamic'' regime $|\p|\ll p_c$ direction fluctuations are described by a non-linear sigma model. It is convenient to use the standard parametrization $\n=(\sig,\pibf)$ where $\sig$ is the component of $\n$ along the direction of order and $\pibf$ a $(N-1)$-component field ($\n^2=\sig^2+\pibf^2=1$). Integrating over $\sig$, one obtains 
\beq
S[\rho',\pibf] = \intr \llbrace \half (\nablabf\rho')^2 + |r_0| \rho'{}^2  + \frac{\rho_0^2}{2} (\nablabf\pibf)^2 \rrbrace 
\label{action2} 
\eeq
for small transverse fluctuations $\pibf$~\cite{note1}. In this limit, we can treat $\pi_i(\r)$ as a variable varying between $-\infty$ and $\infty$. From (\ref{action2}), we deduce the propagator of the $\pibf$ field,
\beq
\mean{\pi_i(\p) \pi_j(-\p)} = \frac{\delta_{i,j}}{\rho_0^2 \p^2} .
\label{Gpi}
\eeq
Again we note that the terms neglected in (\ref{action2}) would only lead to a finite renormalization of the (bare) stiffness $\rho_0^2$ of the non-linear sigma model at sufficiently low temperature. In fact, equation~(\ref{action2}) gives an exact description of the low-energy behavior $|\p|\ll p_c$ if one replaces $\rho_0^2$ by the exact stiffness and $p_c^{-1}=(2|r_0|)^{-1/2}$ by the exact correlation length of the $\rho'$ field. 

We are now in a position to compute the longitudinal and transverse propagators using 
\beq
\begin{split} 
\varphi_l &= \rho \sig = \rho\sqrt{1-\pibf^2} \simeq \rho_0+\rho' - \half \rho_0 \pibf^2  , \\
\varphibf_t &= \rho \pibf \simeq \rho_0 \pibf . 
\end{split}
\label{paraperp} 
\eeq
Since the long-distance physics is governed by transverse fluctuations, we have retained in (\ref{paraperp}) the leading contributions in $\pibf$. Making use of (\ref{Gpi}), one readily obtains 
\beq
G_t(\p) \simeq \rho_0^2 \mean{\pi_i(\p)\pi_i(-\p)} = \frac{1}{\p^2} . 
\label{Gperp}
\eeq
The longitudinal propagator is given by 
\begin{align}
G_l(\r) &= \mean{\rho'(\r)\rho'(0)} + \quarter \rho_0^2 \mean{\pibf(\r)^2 \pibf(0)^2}_c \nonumber \\ 
&= \mean{\rho'(\r)\rho'(0)} + \frac{N-1}{2\rho_0^2} G_t(\r)^2 ,
\label{Gl}
\end{align} 
where $\mean{\cdots}_c$ stands for the connected part of $\mean{\cdots}$. The second line is obtained using Wick's theorem. In Fourier space, this gives 
\beq
G_l(\p) = \frac{1}{\p^2+p_c^2} + \frac{N-1}{2\rho_0^2} \int_\q \frac{1}{\q^2(\p+\q)^2} ,
\label{Gpara} 
\eeq
where the momentum integral is given by (\ref{mint}) for $|\p|\ll\Lambda$ and $d\leq 4$. By comparing the two terms in the rhs of (\ref{Gpara}), we recover the Ginzburg momentum scale (\ref{pG1}). For $|\p|\gg p_G$, the longitudinal propagator $G_{l}(\p)\simeq 1/(\p^2+p_c^2)$ is dominated by amplitude fluctuations and we reproduce the result of the Gaussian approximation. On the other hand, for $|\p|\ll p_G$, $G_l(\p)\sim 1/|\p|^{4-d}$ is dominated by direction fluctuations and diverges for $\p\to 0$. 

The divergence of the longitudinal propagator is a direct consequence of the coupling between longitudinal and transverse fluctuations~\cite{Patasinskij73}. In the long-distance limit, amplitude fluctuations become frozen so that $|\varphibf|=\rho\simeq \rho_0$.  This implies that the longitudinal and transverse components $\varphi_l$ and $\varphibf_t$ cannot be considered independently as in the Gaussian approximation (Sec.~\ref{subsec_phi4_pt}) but satisfy the constraint $\varphi_l^2+\varphibf_t^2\simeq\rho_0^2$. To leading order, $\varphi_l\simeq \rho_0(1-\frac{\pibf^2}{2})^{1/2}$ and $G_l(\r) \sim G_t(\r)^2$ [Eq.~(\ref{Gl})], i.e. $G_l(\p)\sim 1/|\p|^{4-d}$ for $d\leq 4$ (the divergence is logarithmic for $d=4$). 

Equations~(\ref{Gperp}) and (\ref{Gpara}) imply that the self-energies must satisfy 
\beq
\begin{split}
\Sigma_{11}(\p) &= -r_0 - \p^2 + C_1 |\p|^{4-d} , \\
\Sigma_{22}(\p) &= -r_0 + C_2 \p^2 , 
\end{split} 
\eeq
for $\p\to 0$ and $d<4$, i.e.
\beq
\begin{split}
\Sign(\p) &= - r_0 + \frac{C_1}{2} |\p|^{4-d} + \calO(\p^2) , \\
\Sigan(\p) &= \frac{C_1}{2} |\p|^{4-d}+ \calO(\p^2)  .
\end{split}
\label{sigma2a} 
\eeq
For $d=4$, one finds 
\beq
\begin{split}
\Sign(\p) &= - r_0 + \frac{C_1}{\ln(\Lambda/|\p|)} + \calO(\p^2) , \\
\Sigan(\p) &= \frac{C_1}{\ln(\Lambda/|\p|)}+ \calO(\p^2) .
\end{split}
\label{sigma2b} 
\eeq
For $\p=0$, we reproduce the exact results of Sec.~\ref{subsubsec_phi4_sigmaexact}. Equations~(\ref{sigma2a},\ref{sigma2b}) show that $\Sign(\p)$ and $\Sigan(\p)$ contain non-analytic terms that are dominant for $\p\to 0$.

\subsection{Large-$N$ limit} 
\label{subsec_largeN} 

In this section, we show that the previous results for the longitudinal propagator are fully consistent with the large-$N$ limit of the $(\varphibf^2)^2$ theory. Furthermore, the large-$N$ limit enables to compute the longitudinal propagator not only at low temperatures but also in the critical regime near the transition to the high-temperature (disordered) phase. 

To obtain a meaningful large-$N$ limit, we write the coefficient of the $(\varphibf^2)^2$ term in Eq.~(\ref{action}) as $u_0/N$ and take the limit $N\to\infty$ with $u_0$ fixed. Following Ref.~\cite{Zinn_book_2}, we express the partition function as 
\beq
Z = \int \calD[\varphibf,\rho,\lambda] \, e^{ -\intr \left[ \half(\nablabf\varphibf)^2+ \frac{r_0}{2}\rho + \frac{u_0}{4!N} \rho^2 + i \frac{\lambda}{2} (\varphibf^2-\rho) \right] } .
\label{action3} 
\eeq
It can be easily verified that by integrating out $\lambda$ and then $\rho$, one recovers the original action $S[\varphibf]$. If, instead, one first integrates out $\rho$, one obtains
\beq
Z = \int \calD[\varphibf,\lambda] \, e^{ -\intr \left[ \half(\nablabf\varphibf)^2 + i \frac{\lambda}{2} \varphibf^2 \right] + \frac{3N}{2u_0} \intr\,(i\lambda-r_0)^2 } .
\eeq
As in Sec.~\ref{subsec_phi4_hydro}, it is convenient to split the $\varphibf$ field into a $\sig$ field and a $(N-1)$-component field $\pibf$. The integration over the $\pibf$ field gives
\beq
\int\calD[\pibf] \,e^{ -\intr \left[ \half (\nablabf\pibf)^2 + i \frac{\lambda}{2} \pibf^2 \right]} = (\det\, g)^{(N-1)/2} ,
\eeq
where 
\beq
g^{-1}(\r,\r') = - \nablabf^2 \delta (\r-\r') + i\lambda(\r) \delta(\r-\r') 
\eeq
is the inverse propagator of the $\pi_i$ field in the fluctuating $\lamb$ field. We thus obtain the action 
\begin{multline}
S[\sig,\lambda] = \half \intr\left[ (\nablabf\sig)^2 + i \lambda \sig^2 \right] \\ - \frac{3N}{2u_0} \intr\,(i\lambda-r_0)^2 + \frac{N-1}{2} \Tr\ln g^{-1} . 
\label{action4} 
\end{multline} 
In the limit $N\to\infty$, the action becomes proportional to $N$ (this is easily seen by rescaling the $\sig$ field, $\sig\to\sqrt{N}\sig$) and the saddle-point approximation becomes exact. For uniform fields $\sig(\r)=\sig$ and $\lambda(\r)=\lambda$, the action is given by
\beq
\frac{1}{V} S[\sig,\lambda] = \frac{i}{2} \lambda \sig^2 - \frac{3N}{2u_0} (i\lambda-r_0)^2 + \frac{N}{2V} \Tr\ln g^{-1}  
\label{action5} 
\eeq
(we use $N-1\simeq N$ for large $N$), with $g^{-1}(\p)=\p^2+i\lambda$ in Fourier space. From (\ref{action5}), we deduce the saddle-point equations 
\beq
\begin{split}
\sig m^2 &= 0 , \\
\sig^2 &= \frac{6N}{u_0}(m^2-r_0) - N \int_\p \frac{1}{\p^2+m^2} ,
\end{split}
\label{gap_eq}
\eeq
where we use the notation $m^2=i\lambda$ ($i\lambda$ is real at the saddle point). These equations show that the component $\sig$ of the $\varphibf$ field which was singled out plays the role of an order parameter. 

\begin{figure}
\centerline{\includegraphics[width=4.5cm]{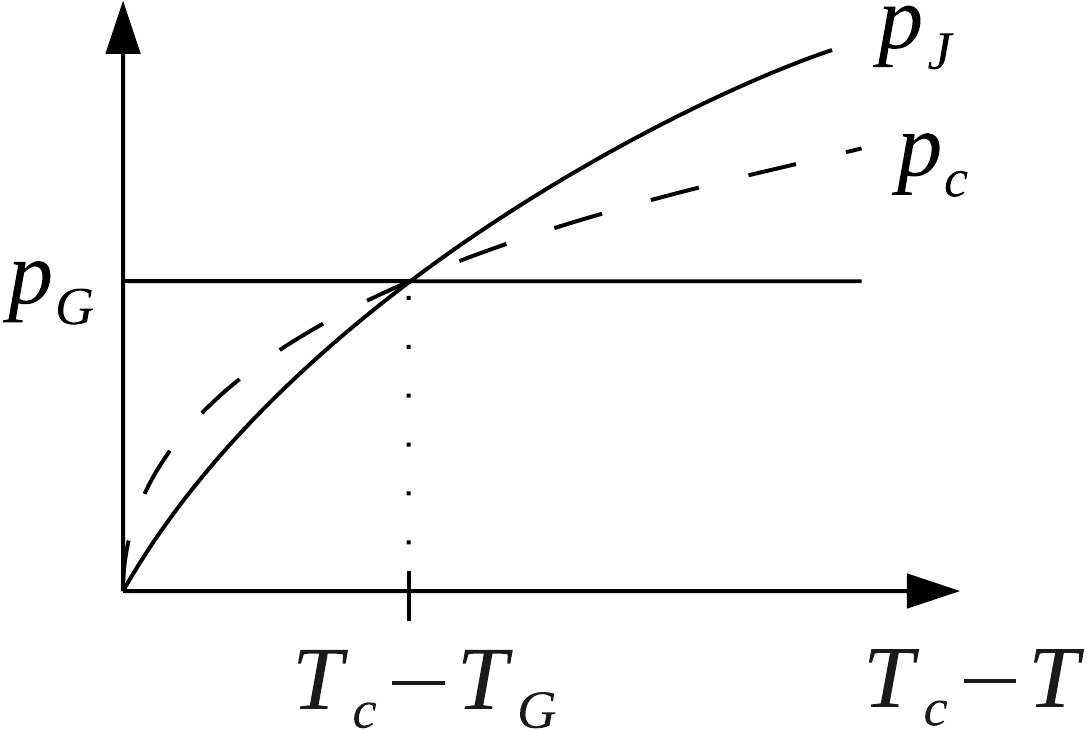}}
\caption{Characteristic momentum scales $p_G$, $p_J$ and $p_c$ vs $T_{c}-T$ for fixed $u_0$ [Eqs.~(\ref{pdef}) with $r_0=\bar r_0(T-T_0)$].}
\label{fig_pGJ} 
\end{figure}

In the low-temperature phase, $\sig$ is non-zero and $m=0$. The propagator $g(\p)=1/\p^2$ is gapless, thus identifying the $\pi_i$ fields as the $N-1$ Goldstone modes associated to the spontaneously broken O($N$) symmetry. From Eq.~(\ref{gap_eq}), we deduce
\beq
\sig^2 = - \frac{6N}{u_0}(r_0-r_{0c}) , 
\eeq
where 
\beq
r_{0c} = - \frac{u_0}{6} \int_\p \frac{1}{\p^2} = - \frac{u_0}{6} \frac{K_d}{d-2} \Lambda^{d-2} 
\eeq
(with $K_d=2^{1-d}\pi^{-d/2}/\Gamma(d/2)$) is the critical value of $r_0$. Since the saddle-point approximation is exact in the large-$N$ limit, the effective action $\Gamma[\sig,\lambda]$ is simply given by the action $S[\sig,\lambda]$ [Eq.~(\ref{action4})]~\cite{note9}. We deduce 
\begin{multline}
\Gamma^{(2)}(\r-\r') = \left( 
\begin{array}{cc} \Gamma^{(2)}_{\sig\sig}(\r-\r') & \Gamma^{(2)}_{\sig\lambda}(\r-\r') \\
\Gamma^{(2)}_{\lambda\sig}(\r-\r') & \Gamma^{(2)}_{\lambda\lambda}(\r-\r') \end{array} \right) \\
= \left( \begin{array}{cc} -\nablabf^2 \delta(\r-\r') & i\sig \delta(\r-\r') \\ i\sig \delta(\r-\r') & \frac{N}{2} \Pi(\r-\r') + \frac{3N}{u_0}\delta(\r-\r') \end{array} \right) ,
\end{multline} 
where 
\beq
\Pi(\r-\r')=g(\r-\r')g(\r'-\r)
\eeq
and we use the notation $\Gamma^{(2)}_{\sig\sig}(\r-\r')=\delta^{(2)}\Gamma/\delta\sig(\r)\delta\sig(\r')$, etc. The two-point vertex $\Gamma^{(2)}$ is computed for the saddle-point values of the fields $\sig$ and $\lambda$. In Fourier space, we obtain
\beq
\Gamma^{(2)}(\p) = \left(  \begin{array}{cc} \p^2 & i\sig \\ i\sig & \frac{N}{2} \Pi(\p) + \frac{3N}{u_0} \end{array} \right) 
\eeq
and the propagator $G=\Gamma^{(2)-1}$ takes the form
\beq
G(\p) = \frac{1}{\det\, \Gamma^{(2)}(\p)} 
\left( \begin{array}{cc} \frac{N}{2} \Pi(\p) + \frac{3N}{u_0} & -i\sig \\ -i\sig & \p^2 \end{array} \right) ,
\eeq
with 
\beq
\det\, \Gamma^{(2)}(\p) = \p^2 \left[ \frac{N}{2} \Pi(\p) + \frac{3N}{u_0} \right] + \sig^2 
\label{gam1}
\eeq
and $\Pi(\p)=\int_\q g(\q)g(\p+\q)$. Equation~(\ref{gam1}), together with the small $\p$ behavior of $\Pi(\p)$ [Eq.~(\ref{mint})], leads us to introduce three characteristic momentum scales,
\beq
\begin{split}
p_G &= \left(\frac{u_0A_d}{6}\right)^{1/(4-d)} , \\
p_J &= \left(\frac{2\sig^2}{NA_d}\right)^{1/(d-2)} = \left[ \frac{12}{u_0A_d}(r_{0c}-r_0)\right]^{1/(d-2)}, \\
p_c &= \left(\frac{u_0\sig^2}{3N}\right)^{1/2} = [2(r_{0c}-r_0)]^{1/2} .
\end{split}
\label{pdef}
\eeq
For simplicity, we discuss only the case $d<4$; equivalent results for $d=4$ are easily deduced. The Josephson length $\xi_J=p_J^{-1}$ -- which separates the critical regime from the Golstone regime (see below)~\cite{Josephson66} -- diverges at the transition with the critical exponent $\nu=1/(d-2)$, which also characterizes the divergence of the correlation length in the high-temperature phase~\cite{Zinn_book_2}. The momentum scales (\ref{pdef}) are not independent since
\beq
p_c^2 = p_G^2 \left(\frac{p_J}{p_G}\right)^{d-2} . 
\eeq
If we vary $r_0$ with $u_0$ fixed, we find that the three characteristic scales (\ref{pdef}) are equal when $T=T_{G}$, where $T_G$ is defined by
\beq
\bar r_0(T_c-T_G) = \half \left(\frac{u_0A_d}{6}\right)^{2/(4-d)} 
\label{critG}
\eeq
(see Fig.~\ref{fig_pGJ}). We have assumed that $r_0=\bar r_0(T-T_0)$ (with $T_0$ the mean-field transition temperature) and used $r_{0c}=\bar r_0(T_c-T_0)$. We recognize in (\ref{critG}) the Ginzburg criterion~\cite{Chaikin_book} so that we can identify $T_G$ with the Ginzburg temperature separating the critical regime near the transition from the non-critical regime at sufficiently low temperatures. 

\begin{figure}
\centerline{\includegraphics[width=6cm]{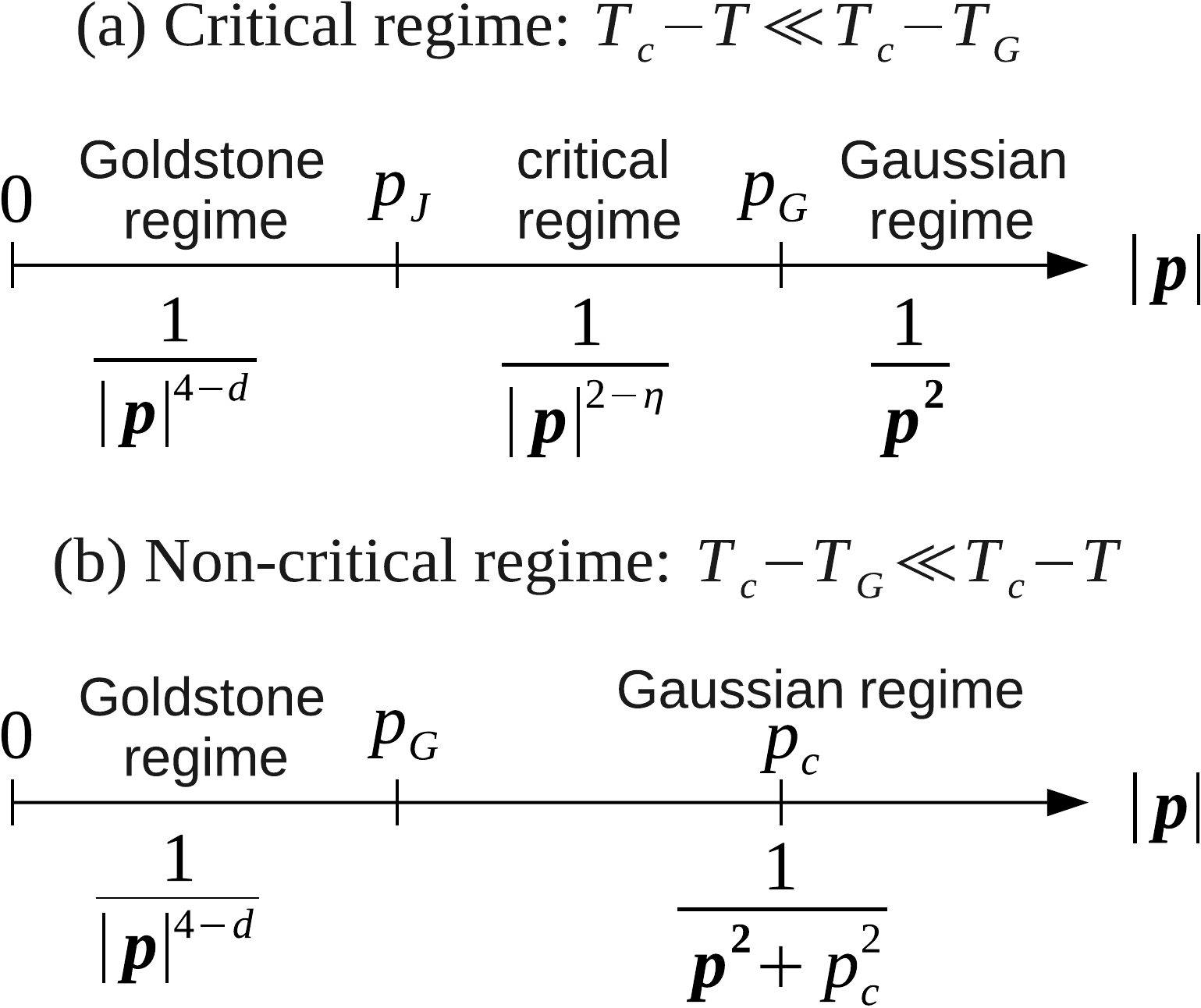}}
\caption{Momentum dependence of the longitudinal correlation function $G_{\sig\sig}(\p)$ in the critical and non-critical regimes of the low-temperature phase as obtained from the large-$N$ limit ($2<d<4$).}
\label{fig_Gsigsig} 
\end{figure}

In the critical regime ($T_c-T\ll T_c-T_G$  or $p_J\ll p_G$), using $p_J\ll p_c\ll p_G$, one finds the longitudinal correlation function 
\beq
G_{\sig\sig}(\p) = \llbrace 
\begin{array}{lcc} 
\dfrac{p_J^{2-d}}{|\p|^{4-d}} & \mbox{if} & |\p| \ll p_J , \\
\dfrac{1}{\p^2} & \mbox{if} & |\p|\gg p_J , 
\end{array}
\right. 
\label{ch_rg:apert11}
\eeq
while in the non-critical regime ($T_c-T_G\ll T_c-T$ or $p_G\ll p_c$),
\beq
G_{\sig\sig}(\p) = \llbrace 
\begin{array}{lcc}
\dfrac{1}{p_c^2} \left(\dfrac{p_G}{|\p|}\right)^{4-d} & \mbox{if} & |\p| \ll p_G , \\
\dfrac{1}{\p^2+p_c^2} & \mbox{if} &  |\p| \gg p_G . 
\end{array}
\right. 
\label{ch_rg:apert12}
\eeq
In the non-critical regime, we recover the results of section~\ref{subsec_phi4_hydro}. We find two characteristic momentum scales ($p_G$ and $p_c$) and two regimes for the behavior of $G_{\sig\sig}(\p)$: i) a Goldstone regime ($|\p|\ll p_G$) characterized by a diverging longitudinal propagator $G_{\sig\sig}(\p)\sim 1/|\p|^{4-d}$, ii) a Gaussian (perturbative) regime ($|\p|\gg p_G$) where $G_{\sig\sig}(\p)\simeq 1/(\p^2+p_c^2)$. The critical regime is characterized by two momentum scales ($p_J$ and $p_G$) and three regimes for the behavior of $G_{\sig\sig}(\p)$: i) a Goldstone regime ($|\p|\ll p_J$) with a diverging longitudinal propagator, ii) a critical regime ($p_J\ll |\p|\ll p_G$) where $G_{\sig\sig}(\p)\sim 1/|\p|^{2-\eta}$ with a vanishing anomalous dimension $\eta$ ($\eta$ is $\calO(1/N)$ in the large-$N$ limit~\cite{Zinn_book_2,note15}), iii) a Gaussian regime ($p_G\ll |\p|$) where $G_{\sig\sig}(\p)\simeq 1/\p^2$. These results are summarized in figure~\ref{fig_Gsigsig}.

\subsection{The non-perturbative RG} 
\label{subsec_phi4_nprg} 

\subsubsection{The average effective action} 

The strategy of the NPRG is to build a family of theories indexed by a momentum scale $k$ such that fluctuations are smoothly taken into account as $k$ is lowered from the microscopic scale $\Lambda$ down to 0~\cite{Berges02,Delamotte07}. This is achieved by adding to the action (\ref{action}) the infrared regulator
\beq
\Delta S_k[\varphibf] = \half \sum_{\p,i} \varphi_i(-\p) R_k(\p) \varphi_i(\p) .
\eeq
The average effective action 
\beq
\Gamma_k[\phibf] = - \ln Z_k[J] + \intr \sum_i J_i \phi_i - \Delta S_k[\phibf] 
\eeq
is defined as a modified Legendre transform of $-\ln Z_k[J]$ which includes the subtraction of $\Delta S_k[\phibf]$. Here $J_i$ is an external source which couples linearly to the $\varphi_i$ field and $\phibf(\r)=\mean{\varphibf(\r)}_J$. The cutoff function $R_k$ is chosen such that at the microscopic scale $\Lambda$ it suppresses all fluctuations, so that the mean-field approximation $\Gamma_\Lambda[\phibf]=S[\phibf]$ becomes exact. The effective action of the original model (\ref{action}) is given by $\Gamma_{k=0}$ provided that $R_{k=0}$ vanishes. For a generic value of $k$, 
the cutoff function $R_k(\p)$ suppresses fluctuations with momentum $|\p|\lesssim k$ but leaves unaffected those with $|\p|\gtrsim k$. The variation of the average effective action with $k$ is governed by Wetterich's equation~\cite{Wetterich93} 
\beq
\dt \Gamma_k[\phibf] = \half \Tr\llbrace \dot R_k\left(\Gamma^{(2)}_k[\phibf] + R_k\right)^{-1} \rrbrace ,
\label{rgeq}
\eeq
where $t=\ln(k/\Lambda)$ and $\dot R_k=\dt R_k$. $\Gamma^{(2)}_k[\phibf]$ denotes the second-order functional derivative of $\Gamma_k[\phibf]$. In Fourier space, the trace involves a sum over momenta as well as the internal index of the $\phibf$ field.  

Because of the regulator term $\Delta S_k$, the vertices $\Gamma^{(n)}_{k,i_1\cdots i_n}(\p_1,\cdots,\p_n)$ are smooth functions of momenta and can be expanded in powers of $\p_i^2/k^2$. Thus if we are interested only in the long distance physics, we can use a derivative expansion of the average effective action~\cite{Berges02,Delamotte07}. In the following, we consider the ansatz
\beq
\Gamma_k[\phibf] = \intr \llbrace \frac{Z_k}{2} (\nablabf\phibf)^2 + U_k(\rho) \rrbrace .
\label{ansatz1}
\eeq
Because of the O($N$) symmetry, the effective potential $U_k(\rho)$ must be a function of the O($N$) invariant $\rho=\phibf^2/2$. To further simplify the analysis, we expand $U_k(\rho)$ about its minimum $\rho_{0,k}$, 
\beq
U_k(\rho) = U_k(\rho_{0,k}) + \frac{\lambda_k}{2}(\rho-\rho_{0,k})^2 . 
\label{ansatz2}
\eeq
We consider only the ordered phase where $\rho_{0,k}>0$. In a broken symmetry state with order parameter $\phibf=(\sqrt{2\rho_{0,k}},0,\cdots,0)$, the two-point vertex is given by 
\beq
\Gamma^{(2)}_{k,ii}(\p) = \llbrace 
\begin{array}{lcc}
Z_k\p^2 + 2\lamb_k\rho_{0,k} & \mbox{if} & i=1 , \\ 
Z_k\p^2 & \mbox{if} & i \neq 1 .
\end{array}
\right. 
\eeq
By inverting $\Gamma^{(2)}_k$, we obtain the longitudinal and transverse parts of the propagator, 
\beq
\begin{split} 
G_{k,l}(\p) &= \frac{1}{Z_k\p^2 + 2\lamb_k\rho_{0,k}} , \\ 
G_{k,t}(\p) &= \frac{1}{Z_k\p^2} . 
\end{split}
\label{G3}
\eeq
Since these expressions are obtained from a derivative expansion of the average effective action, they are valid only in the limit $|\p|\ll k$. In practice however, one can retrieve the momentum dependence of $G_{k=0}(\p)$ at finite $\p$ by stopping the RG flow at $k\sim|\p|$, i.e. $G_{k=0}(\p)\simeq G_{k\sim|\p|}(\p)$, where $G_{k\sim|\p|}(\p)$ can be approximated by the result of the derivative expansion. It is possible to obtain the full momentum dependence of the correlation functions in a more rigorous and precise way, within the so-called Blaizot-Mendez-Weschbor scheme~\cite{Blaizot06,Benitez09,Ledowski04}, but this requires a much more involved numerical analysis of the RG equations. 

The transverse propagator $G_{k,t}(\p)$  is gapless [Eq.~(\ref{G3})], in agreement with Goldstone's theorem, which is a mere consequence of the O($N$) symmetry of the average effective action (\ref{ansatz1}). On the other hand, the divergence of the longitudinal susceptibility obtained in the previous sections suggests that $\lamb_k\to 0$ for $k\to 0$ ($\lim_{k\to 0}\rho_{0,k}>0$ in the ordered phase). We shall see that this is indeed the result obtained from the RG equations. 

\subsubsection{RG flows} 
\label{subsubsec_phi4_rgeq} 

It is convenient to work with the dimensionless variables 
\beq
\begin{split}
\tilde\rho_{0,k} &= Z_k k^{2-d}\rho_{0,k}, \\ 
\tilde\lambda_k &= Z_k^{-2} k^{d-4} \lambda_k . 
\end{split} 
\eeq
The flow equations for $\tilde\rho_{0,k}$, $\tilde\lambda_k$ and $Z_k$ are obtained by inserting the ansatz (\ref{ansatz1},\ref{ansatz2}) into the RG equation (\ref{rgeq}). The calculation is standard~\cite{Berges02,Delamotte07} and we only quote the final result,
\beq
\begin{split}
\dt \tilde\rho_{0,k} ={}& (2-d-\eta_k) \tilde\rho_{0,k} - \frac{3}{2} \tilde I_{k,l} - \frac{N-1}{2} \tilde I_{k,t} , \\
\dt \tilde\lamb_k ={}& (d-4+2\eta_k) \tilde\lamb_k \\ & - \tilde\lamb_k^2 [9 \tilde J_{k,ll}(0) + (N-1) \tilde J_{k,tt}(0)] , \\
\eta_k =&{} 2\tilde\lamb_k^2 \tilde\rho_{0,k} [\tilde J'_{k,lt}(0) + \tilde J'_{k,tl}(0) ] ,
\end{split}
\label{rgeq1} 
\eeq
where $\eta_k=-\dt\ln Z_k$ denotes the running anomalous dimension. With the cutoff function $R_k(\p)=Z_k (\p^2-k^2)\Theta(\p^2-k^2)$~\cite{Litim00} ($\Theta(x)$ is the step function), the threshold functions appearing in (\ref{rgeq1}) can be calculated analytically (see Appendix \ref{sec_threshold}). 

\begin{figure}
\centerline{\includegraphics[width=6cm,clip]{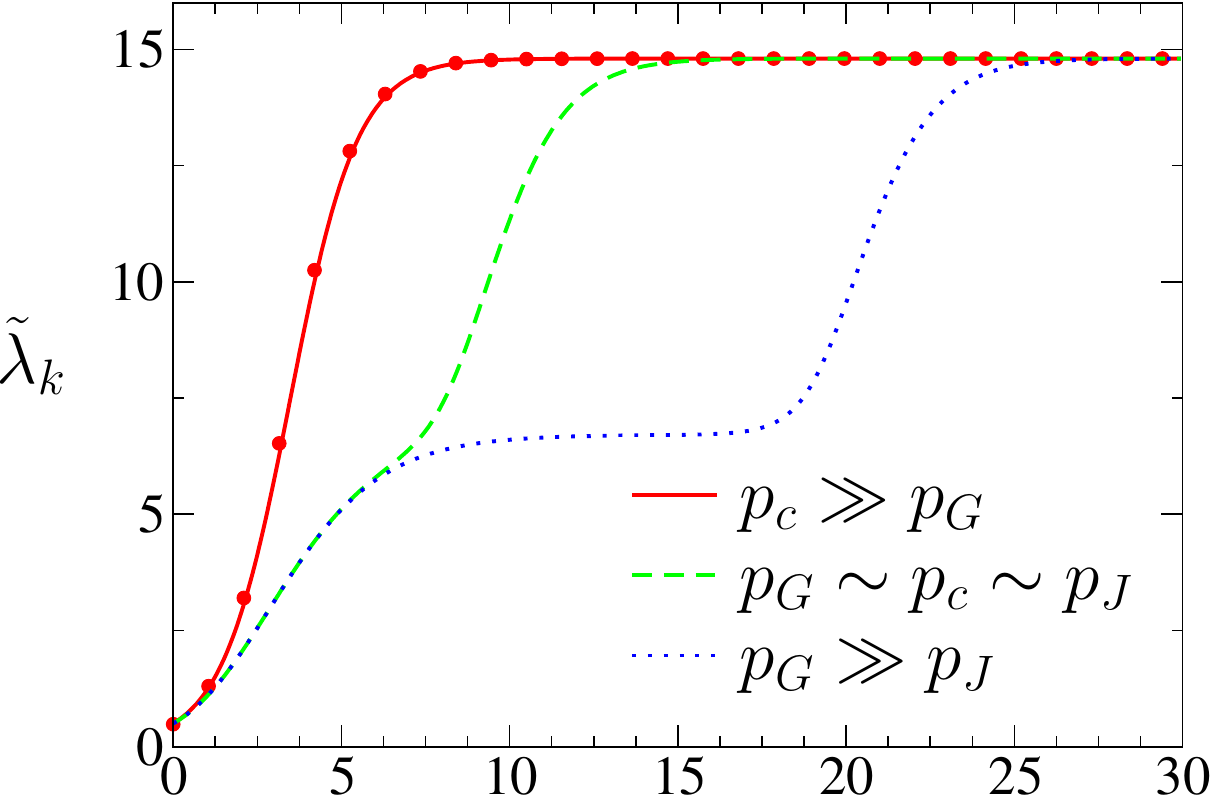}} 
\vspace{0.3cm}
\centerline{\includegraphics[width=6cm,clip]{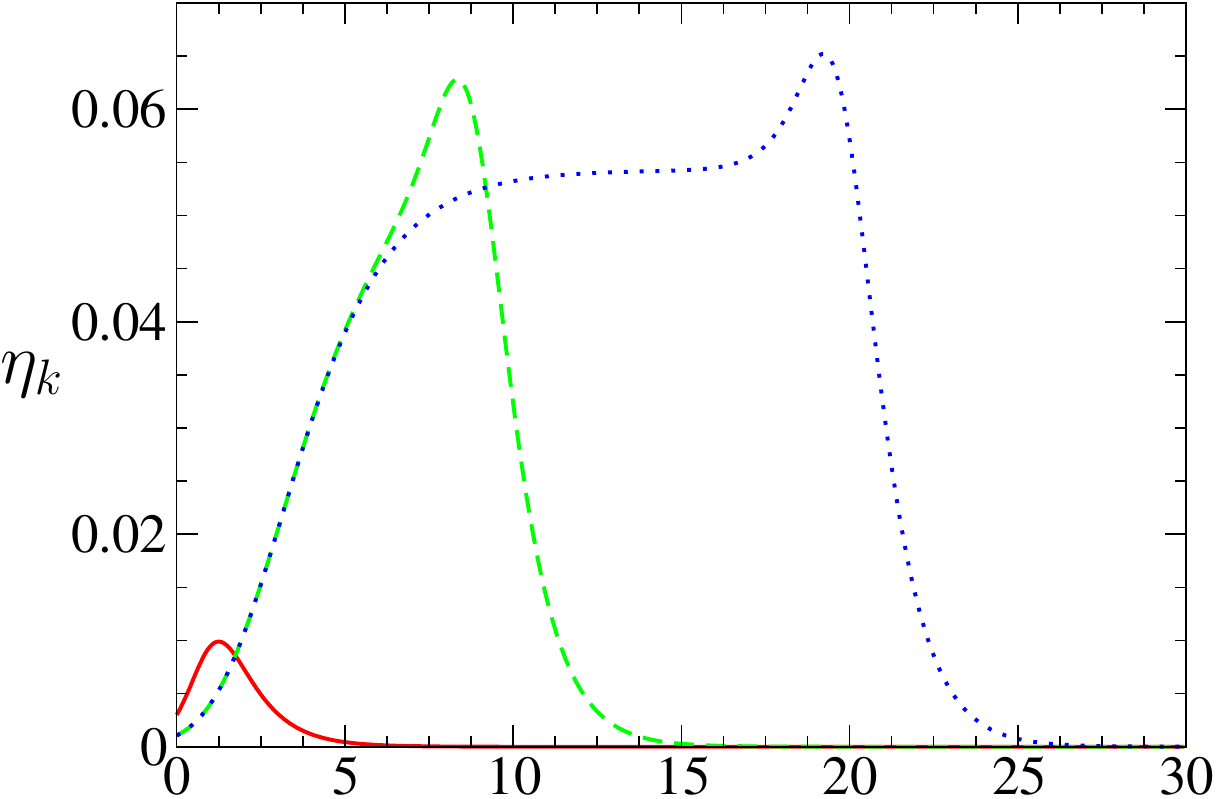}} 
\vspace{0.3cm}
\centerline{\includegraphics[width=6cm,clip]{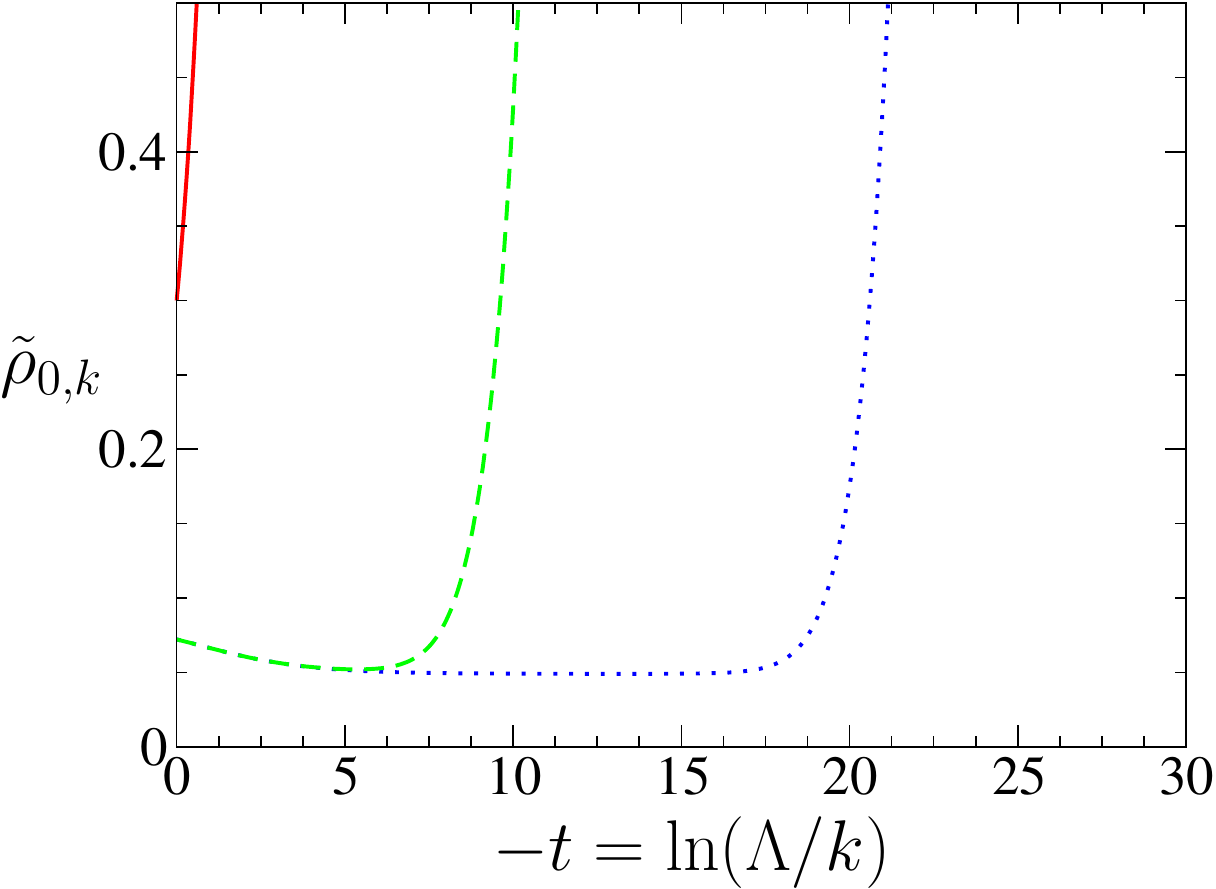}}
\caption{(Color online) $\tilde\lamb_k$, $\eta_k$ and $\tilde\rho_{0,k}$ vs $-t=\ln(\Lambda/k)$ for $d=3$, $N=3$, $\Lambda=1$ and $\lamb_{k=0}=0.5$. The (red) solid line corresponds to $\rho_{0,k=0}=0.3$ ($p_c\gg p_G$) and the (red) dots are obtained from the analytic solution (\ref{lambk}). The (green) dashed line corresponds to $\rho_{0,k=0}=0.072147$ ($p_c\sim p_G\sim p_J$) and the (blue) dotted one to $\rho_{0,k=0}=0.072146123$ ($p_G\gg p_J$).} 
\label{fig_flow}  
\end{figure}

In Fig.~\ref{fig_flow} we show $\tilde\lamb_k$, $\eta_k$ and $\tilde\rho_{0,k}$ vs $-t=\ln(\Lambda/k)$ for $d=3$ and $N=3$. We fix $\lamb_{k=0}=u_0/3$ and vary $r_0$ (i.e. $\rho_{0,k=0}=-3r_0/u_0$). When the system is in the ordered phase away from the critical regime (red solid lines in Fig.~\ref{fig_flow}), i.e. $p_c\gg p_G$, we see a crossover for $k\sim p_G$ ($t_G=\ln(p_G/\Lambda)\simeq -4$) from the Gaussian regime to the Goldstone regime characterized by $\tilde\lamb_k\simeq\tilde\lamb^*$, $\eta_k=0$ and $\tilde\rho_{0,k}\sim k^{-1}$ (i.e. $\rho_{0,k}\simeq \rho_0^*=\lim_{k\to 0}\rho_{0,k}$). Since $\tilde\lamb_k\simeq\tilde\lamb^*$ and $\eta_k=0$ imply $\lamb_k\sim k$, we find that the longitudinal susceptibility $G_{k,l}(\p)=1/2\lamb_k\rho_{0,k}\sim 1/k$ diverges when $k\to 0$. Identifying $k$ with $|\p|$ to extract the momentum dependence (as explained above), we recover the singular behavior $G_{k=0,l}(\p)\sim 1/|\p|$ in three dimensions. More generally, for an arbitrary dimension, one finds $\lamb_k\sim k^\eps \tilde\lamb^*$ and $G_{k,l}(\p)\sim 1/k^\eps\equiv 1/|\p|^\eps$ with $\eps=4-d$. Thus in the RG approach, the divergence of the longitudinal susceptibility is a consequence of the existence of a fixed point for the dimensionless coupling constant $\tilde\lamb_k$. 

When the system is in the critical regime of the ordered phase (blue dotted lines in Fig.~\ref{fig_flow}), i.e. $p_G\gg p_J$, there is a first crossover from the Gaussian regime to the critical regime for $k\sim p_G$ followed by a second crossover to the Goldstone regime for $k\sim p_J$. In the critical regime $p_G\gg k\gg p_J$, $\tilde\lamb_k\simeq \tilde\lamb_{\rm cr}^*$, $\eta_k\simeq \eta^*$ and $\tilde\rho_{0,k}\simeq \tilde\rho_0^*$ are nearly equal to their values at the critical point between the ordered and disordered phases~\cite{Tetradis94,note3}. This gives $G_{k,t}(\p)\simeq G_{k,l}(\p) \sim 1/k^{-\eta^*}\p^2$, i.e.  $G_{k=0,t}(\p)\simeq G_{k=0,l}(\p) \sim 1/|\p|^{2-\eta^*}$ if we identify $k$ with $|\p|$.

\subsubsection{Analytical solution in the low-temperature phase} 

In the low-temperature phase (away from the critical regime, i.e. when $p_c\gg p_G$), it is possible to obtain an analytical solution of the flow equations for $k\ll p_c$. In this limit, the RG flow is dominated by the Goldstone modes and the contribution of the longitudinal mode can be omitted. This amounts to ignoring $\tilde J_{k,ll}(0)$, $\tilde J'_{k,lt}(0)$ and $\tilde J'_{k,tl}(0)$ in Eqs.~(\ref{rgeq1}), which is justified by noting that $\tilde\lambda_k\tilde\rho_{0,k}$ becomes very large for $k\ll p_c$ ($\tilde\lambda_k\tilde\rho_{0,k}\sim k^{2-d}$ for $k\to 0$), where the hydrodynamic scale $p_c$ is defined by $2\tilde\lambda_{p_c}\tilde\rho_{0,p_c}\sim 1$. This gives $\eta_k=0$ and 
\beq
\dt\tilde\lamb_k = - \eps \tilde\lamb_k + 8 \frac{v_d}{d} (N-1) \tilde\lamb_k^2 , 
\label{rgeq2} 
\eeq
where $v_d=[2^{d+1}\pi^{d/2}\Gamma(d/2)]^{-1}$. We have used the expression of the threshold functions given in Appendix~\ref{sec_threshold}. Equation~(\ref{rgeq2}) should be solved with the boundary condition $\tilde\lamb_k=\tilde\lamb_c$ for $k=\Lambda_0\simeq p_c$. For $d<4$, we then find
\begin{align}
\tilde\lamb_k &= \frac{\eps \tilde\lamb_c p_c^\eps}{\eps k^\eps+ 8\frac{v_d}{d}(N-1) \tilde\lamb_c(p_c^\eps-k^\eps)} \nonumber \\ 
&\simeq \frac{\eps \tilde\lamb_c p_c^\eps}{\eps k^\eps + 8\frac{v_d}{d}(N-1) \tilde\lamb_c p_c^\eps}
\end{align}
for $k\ll p_c$. The last expression can be rewritten in the more insightful form 
\beq
\tilde\lamb_k = \frac{\tilde\lamb^*}{1+(k/p_G)^\eps} , 
\label{lambk} 
\eeq
where
\beq
\tilde\lamb^* = \lim_{k\to 0} \tilde\lamb_k = \frac{\eps d}{8v_d(N-1)} 
\label{lambfp} 
\eeq
and
\begin{align}
p_G &= \biggl[(N-1) \frac{8v_dp_c^\eps\tilde\lamb_c}{d\eps}\biggr]^{1/\eps} \nonumber \\ 
&= \biggl[(N-1) \frac{8v_d\lamb_c}{d\eps Z_{p_c}^2}\biggr]^{1/\eps}  .
\end{align}  
Equation~(\ref{lambk}) is in remarkable agreement with the numerical solution of the flow equations (\ref{rgeq1}) (Fig.~\ref{fig_flow}). In the weak-coupling limit $p_G\ll p_c$, we can ignore the renormalization of $Z_k$ as well as that of $\lamb_k$ between $k=\Lambda$ and $k=p_c$, and approximate $Z_{p_c}\simeq 1$ and $\lambda_c\simeq \lamb_{k=\Lambda}=u_0/3$. We then recover the expression 
\beq
p_G \simeq  \left[(N-1) \frac{8v_du_0}{3d\eps}\right]^{1/\eps}
\label{pG2}
\eeq
of the Ginzburg momentum scale obtained in previous sections. A similar analysis can be made for the case $d=4$.

\section{Interacting bosons}
\label{sec_bosons} 

We consider interacting bosons at zero temperature with the (Euclidean) action
\begin{equation}
S = \int dx \left[ \psi^*\left(\dtau-\mu - \frac{\nablabf^2}{2m}
  \right) \psi + \frac{g}{2} (\psi^*\psi)^2 \right] ,
\end{equation}
where $\psi(x)$ is a bosonic (complex) field, $x=(\r,\tau)$, and $\int dx=\inttau \int d^dr$. $\tau\in [0,\beta]$ is an imaginary time, $\beta\to\infty$ the inverse temperature, and $\mu$ denotes the
chemical potential. The interaction is assumed to be local in space and the
model is regularized by a momentum cutoff $\Lambda$. We consider a space dimension $d>1$.

Introducing the two-component field 
\begin{equation}
\Psi(p) = \left( \begin{array}{c} \psi(p) \\ \psi^*(-p)  \end{array} \right) , \quad 
\Psi^\dagger(p) = \bigl( \psi^*(p), \psi(-p) \bigr) 
\end{equation}
(with $p=(\p,i\w)$ and $\w$ a Matsubara frequency), the one-particle (connected) propagator becomes a $2\times 2$ matrix whose inverse in Fourier space is given by
\begin{equation}
\left( 
\begin{array}{cc} i\w + \mu -\eps_\p -\Sign(p) & - \Sigan(p) \\
 -\Sigan^*(p) & -i\w + \mu -\eps_\p -\Sign(-p)
\end{array}
\right) ,
\label{propa}
\end{equation}
where $\Sign$ and $\Sigan$ are the normal and anomalous self-energies, respectively, and $\eps_\p=\p^2/2m$. If we choose the order parameter $\mean{\psi(x)}=\sqrt{n_0}$ to be real (with $n_0$ the condensate density), then the anomalous self-energy $\Sigan(p)$ is real. 

To make contact with the classical $(\varphibf^2)^2$ theory with O($N$) symmetry studied in Sec.~\ref{sec_phi4}, it is convenient to write the boson field
\beq
\psi(x) = \frac{1}{\sqrt{2}} [\psi_1(x) + i\psi_2(x)] 
\eeq
in terms of two real fields $\psi_1$ and $\psi_2$ and consider the (connected) propagator $G_{ij}(x,x')=\mean{\psi_i(x)\psi_j(x')}_c$. The inverse propagator $G^{-1}_{ij}(p)$ reads
\beq
\left(
\begin{array}{cc}
\eps_\p-\mu+\Sigma_{11}(p) & \w + \Sigma_{12}(p) \\ 
-\w + \Sigma_{21}(p) & \eps_\p-\mu + \Sigma_{22}(p) 
\end{array}
\right) , 
\label{Ga}
\eeq
where 
\beq
\begin{split}
\Sigma_{11}(p) &= \half [\Sign(p)+\Sign(-p)] + \Sigan(p) , \\ 
\Sigma_{22}(p) &= \half [\Sign(p)+\Sign(-p)] - \Sigan(p) , \\
\Sigma_{12}(p) &= \frac{i}{2} [\Sign(p)-\Sign(-p)] , \\
\Sigma_{21}(p) &= -\frac{i}{2} [\Sign(p)-\Sign(-p)] ,
\end{split}
\label{Gb}
\eeq
when $\Sigan(p)$ is real. 

\subsection{Perturbation theory and infrared divergences}
\label{subsec_bosons_pt}

\subsubsection{Bogoliubov's theory} 
\label{subsubsec_bog}

The Bogoliubov approximation is a Gaussian fluctuation theory about the saddle point solution $\psi(x)=\sqrt{n_0}=\sqrt{\mu/g}$ (i.e. $\psi_1(x)=\sqrt{2n_0}$ and $\psi_2(x)=0$). It is equivalent to a zero-loop calculation of the self-energies,
\begin{equation}
\Sign^{(0)}(p) = 2gn_0 , \quad
\Sigan^{(0)}(p) = gn_0 ,
\end{equation}
or, equivalently, 
\beq
\Sigma^{(0)}_{11}(p) = 3gn_0, \quad \Sigma^{(0)}_{22}(p)=gn_0, \quad \Sigma^{(0)}_{12}(p)=0 .
\eeq
This yields the (connected) propagators 
\beq
\begin{split}
G_{\rm n}^{(0)}(p) &= -\mean{\psi(p)\psi^*(p)}_c = \frac{-i\w-\eps_\p-gn_0}{\w^2+E_\p^2} ,\\ 
G_{\rm an}^{(0)}(p) &= -\mean{\psi(p)\psi(-p)}_c = \frac{gn_0}{\w^2+E_\p^2} ,
\end{split}
\eeq
where $E_\p=[\eps_\p(\eps_\p+2gn_0)]^{1/2}$ is the Bogoliubov quasi-particle excitation energy. When $|\p|$ is larger than the healing momentum $p_c=(2gmn_0)^{1/2}$, the spectrum $E_\p\simeq \eps_\p+gn_0$ is particle-like, whereas it becomes sound-like for $|\p|\ll p_c=\sqrt{2}mc$ with a velocity $c=\sqrt{gn_0/m}$. In the weak-coupling limit, $n_0\simeq \bar n$ ($\bar n$ is the mean boson density) and $p_c$ can equivalently be defined as $p_c=(2gm\bar n)^{1/2}$. In the hydrodynamic regime $|\p|\ll p_c$,
\beq
\begin{split}
G_{11}^{(0)}(p) &= \frac{\eps_\p}{\w^2+c^2 \p^2} , \\ 
G_{22}^{(0)}(p) &= \frac{2gn_0}{\w^2+c^2 \p^2} , \\ 
G_{12}^{(0)}(p) &= -\frac{\w}{\w^2+c^2 \p^2} .
\end{split}
\label{Ghydro}
\eeq 
Note that in the Bogoliubov approximation, the occurrence of a linear spectrum at low energy (which implies superfluidity according to Landau's criterion), is due to $\Sigan(0)$ being nonzero. 

\subsubsection{Infrared divergences and the Ginzburg scale}

Let us now consider the one-loop correction $\Sigma^{(1)}$ to the Bogoliubov result $\Sigma^{(0)}$. For $d\leq 3$, the second diagram of Fig.~\ref{fig_self_1loop} gives a divergent contribution when the two internal lines correspond to transverse fluctuations, which is possible only for $\Sigma_{11}$. Thus $\Sigma_{22}$ is finite at the one-loop level and the normal and anomalous self-energies exhibit the same divergence, 
\beq
\Sign^{(1)}(p) \simeq \Sigan^{(1)}(p) \simeq - \half g^2 n_0 \int_q G_{22}^{(0)}(q) G_{22}^{(0)}(p+q) ,
\label{sigma3}
\eeq
where we use the notation $q=(\q,i\w')$ and $\int_q=\intinf\frac{d\w'}{2\pi} \int_\q$. For small $p$, the main contribution to the integral in (\ref{sigma3}) comes from momenta $|\q|\lesssim p_c$ and frequencies $|\w'|\lesssim cp_c$, so that we can use (\ref{Ghydro}) and obtain 
\beq
\Sign^{(1)}(p) \simeq \Sigan^{(1)}(p) \simeq - 2 \frac{g^4 n_0^3}{c^3} \int_\Q \frac{1}{\Q^2(\Q+\P)^2} ,
\label{sigma4}
\eeq
where $\Q=(\q,\w'/c)$ and $\P=(\p,\w/c)$ are $(d+1)$-dimensional vectors. The momentum integral in (\ref{sigma4}) is restricted by $|\Q|\lesssim p_c$ and is given by (\ref{mint}), with $\Lambda$ replaced by $p_c$, $d$ by $d+1$ and $|\p|$ by $(\p^2+\w^2/c^2)^{1/2}$. We can estimate the characteristic (Ginzburg) momentum scale $p_G$ below which the Bogoliubov approximation breaks down from the condition $|\Sign^{(1)}(p)| \sim \Sign^{(0)}(p)$ or  $|\Sigan^{(1)}(p)| \sim \Sigan^{(0)}(p)$ for $|\p|=p_G$ and $|\w|=cp_G$,
\begin{equation}
p_G \sim \left\lbrace 
\begin{array}{lcc}
(A_{d+1} gm p_c)^{1/(3-d)} & \mbox{if} & d<3 , \\
p_c \exp\left( - \frac{1}{A_4 gmp_c}\right) & \mbox{if} & d=3 .
\end{array}
\right.
\label{pG_est}
\end{equation}
This result can be rewritten as
\begin{equation}
p_G \sim \left\lbrace 
\begin{array}{lcc}
p_c (A_{d+1} \tilde g^{d/2})^{1/(3-d)} & \mbox{if} & d<3 , \\
p_c \exp\left( - \frac{1}{A_4 \sqrt{2} \tilde g^{3/2}}\right) & \mbox{if} & d=3 ,
\end{array}
\right.
\end{equation}
where 
\beq
\tilde g = gm \bar n^{1-2/d} \sim \left(\frac{p_c}{\bar n^{1/d}}\right)^2
\eeq
is the dimensionless coupling constant obtained by comparing the mean interaction energy per particle $g\bar n$ to the typical kinetic energy $1/m\bar r^2$ where $\bar r\sim \bar n^{-1/d}$ is the mean distance between particles~\cite{Petrov04}. A superfluid is weakly correlated if $\tilde g\ll 1$, i.e. $p_G\ll p_c\ll \bar n^{1/d}$ (the characteristic momentum scale $\bar n^{1/d}$ does however not play any role in the weak-coupling limit)~\cite{Capogrosso10}. In this case, the Bogoliubov theory applies to a large part of the spectrum where the dispersion is linear (i.e. $|\p|\lesssim p_c$) and breaks down only at very small momenta  $|\p|\lesssim p_G\ll p_c$. In the next sections, we shall see that the weakly-correlated superfluid bears many similarities with the ordered phase of the classical O($N$) model away from the critical regime. When the dimensionless coupling $\tilde g$ becomes of order unity, the three characteristic momentum scales $p_G\sim p_c\sim \bar n^{1/d}$ become of the same order. The momentum range $[p_G,p_c]$ where the linear spectrum can be described by the Bogoliubov theory is then suppressed. We expect the strong-coupling regime $\tilde g\gg 1$ to be governed by a single characteristic momentum scale, namely $\bar n^{1/d}$.

\subsubsection{Vanishing of the anomalous self-energy} 
\label{subsubsec_bosons_sigmaexact}

The exact values of $\Sign(p=0)$ and $\Sigan(p=0)$ can be obtained using the U(1) symmetry of the action, i.e. the invariance under the field transformation $\psi(x)\to e^{i\theta}\psi(x)$ and $\psi^*(x)\to e^{-i\theta}\psi^*(x)$~\cite{Nepomnyashchii75,note2}. The derivation is similar to that of Sec.~\ref{subsubsec_phi4_sigmaexact}. Let us consider the effective action 
\beq
\Gamma[\phi] = -\ln Z[J_1,J_2] + \int dx [J_1\phi_1 + J_2\phi_2] ,
\eeq
where $J_i$ is an external source which couples linearly to the boson field $\psi_i$, and $\phi_i(x)=\mean{\psi_i(x)}_J$ the superfluid order parameter. The U(1) symmetry of the action implies that $\Gamma[\phi]$ is invariant under a uniform rotation of the vector field $(\phi_1(x),\phi_2(x))^T$. For an infinitesimal rotation angle $\theta$, this yields
\beq
\int dx \sum_{i,j} \frac{\delta\Gamma[\phi]}{\delta\phi_i(x)} \eps_{ij} \phi_j(x) = 0 ,  
\label{wi3}
\eeq
where $\eps_{ij}$ is the totally antisymmetric tensor. Taking the functional derivative $\delta/\delta\phi_l(y)$ and setting $\phi_i(x)=\delta_{i,1} \sqrt{2n_0}$ leads to
\beq	
\Gamma^{(2)}_{2l}(p=0) = 0. 
\label{wi2}
\eeq
For $l=2$, equation~(\ref{wi2}) yields the Hugenholtz-Pines theorem~\cite{Hugenholtz59} 
\beq
\Gamma^{(2)}_{22}(p=0) = \Sign(p=0)-\Sigan(p=0)-\mu = 0. 
\label{hpth} 
\eeq
If we now take the second-order functional derivative $\delta^{(2)}/\delta\phi_l(y)\delta\phi_m(z)$ of (\ref{wi3}) and set $\phi_i(x)=\delta_{i,1} \sqrt{2n_0}$, we obtain the Ward identity
\begin{align}
&\sum_i \Gamma^{(2)}_{im}(y,z) \eps_{il} + \sum_i \Gamma^{(2)}_{il}(z,y) \eps_{im} \nonumber \\ &
- \sqrt{2n_0} \int dx\, \Gamma^{(3)}_{2lm}(x,y,z) = 0. 
\end{align}
Integrating over $y$ and $z$ and setting $l=2$ and $m=1$, we deduce (in Fourier space)
\beq
\Gamma^{(3)}_{122}(0,0,0) = \frac{1}{\sqrt{\beta V}} \frac{\Gamma^{(2)}_{11}(0,0)}{\sqrt{2n_0}} ,
\label{wi4}
\eeq
where we have used (\ref{hpth}). 

The self-energy $\Sigma_{11}$ can be written as 
\begin{align}
\Sigma_{11}(p) ={}& \tilde\Sigma_{11}(p) - g\sqrt{\frac{n_0}{2\beta V}} \sum_q G_{22}(q) G_{22}(p+q) \nonumber \\ & \times \Gamma^{(3)}_{122}(-p,-q,p+q) , 
\end{align} 
where $\tilde\Sigma_{11}(p)$ denotes the regular part of the self-energy (i.e. the part that does not contain pairs of lines corresponding to $G_{22}G_{22}$). If we assume that the transverse propagator $G_{22}(q)\sim 1/(\w^2+c^2\q^2)$ at low energies (this result will be shown in the following sections), the integral $\int_q G_{22}(q)^2$ is infrared divergent for $d\leq 3$. To obtain a finite self-energy $\Sigma_{11}(p=0)$, one must require that $\Gamma^{(3)}_{122}(0,0,0)=0$. The Ward identity (\ref{wi4}) then implies $\Gamma^{(2)}_{11}(p=0)=0$ and in turn 
\beq
\begin{split}
\Sign(p=0) &= \mu + \half \left[\Gamma^{(2)}_{11}(p=0) + \Gamma^{(2)}_{22}(p=0) \right] = \mu \\
\Sigan(p=0) &= \half \left[\Gamma^{(2)}_{11}(p=0) - \Gamma^{(2)}_{22}(p=0) \right] = 0 .
\end{split}
\label{sigma6} 
\eeq
The vanishing of the anomalous self-energy $\Sigan(p=0)$ was first proven by Nepomnyashchii and Nepomnyashchii~\cite{Nepomnyashchii75}. To reconcile this result with the existence of a sound mode with linear dispersion, the self-energies $\Sign(p)$ and $\Sigan(p)$ must necessarily contain non-analytic terms in the limit $p\to 0$ (Sec.~\ref{subsubsec_bosons_self}).

\subsection{Hydrodynamic approach} 
\label{subsec_bosons_hydro} 

It was realized by Popov that the phase-density representation of the boson field $\psi=\sqrt{n}e^{i\theta}$ leads to a theory free of infrared divergences~\cite{Popov72,Popov_book_2,Nepomnyashchii83}. Popov's theory bears some similarities with the analysis of the $(\varphibf^2)^2$ theory based on the amplitude-direction representation (Sec.~\ref{subsec_phi4_hydro}). In this section, we show how the phase-density representation can be used to obtain the infrared behavior of the propagators $G_{\rm n}(p)$ and $G_{\rm an}(p)$ without encountering infrared divergences~\cite{Popov79}. Our approach is similar to that of Popov (with some technical differences in Sec.~\ref{subsubsec_hydro}).

\subsubsection{Perturbative approach} 

In terms of the density and phase fields, the action reads
\beq
S[n,\theta] = \int dx \left[ in \dot\theta + \frac{n}{2m}(\nablabf\theta)^2 + \frac{(\nablabf n)^2}{8mn} - \mu n + \frac{g}{2} n^2 \right] .
\label{action6}
\eeq
At the saddle-point level, $n(x)=\bar n=\mu/g$. Expanding the action to second order in $\delta n=n-\bar n$, $\dot\theta$ and $\nablabf\theta$, we obtain 
\beq
S[\delta n,\theta] = \int dx \left[ i\delta n \dot\theta + \frac{\bar n}{2m}(\nablabf\theta)^2 + \frac{(\nablabf n)^2}{8m\bar n} + \frac{g}{2} (\delta n)^2 \right] .
\label{actionh} 
\eeq
The higher-order terms can be taken into account within perturbation theory and only lead to finite corrections of the coefficients of the hydrodynamic action (\ref{actionh})~\cite{Popov_book_2}. 

We deduce the correlation functions of the hydrodynamic variables, 
\beq
\begin{split}
G_{nn}(p) &= \mean{\delta n(p)\delta n(-p)} = \frac{\bar n}{m} \frac{\p^2}{\w^2+E_\p^2} , \\ 
G_{n\theta}(p) &= \mean{\delta n(p)\theta(-p)} = -\frac{\w}{\w^2+E_\p^2} , \\
G_{\theta\theta}(p) &= \mean{\theta(p)\theta(-p)} = \frac{\frac{\p^2}{4m\bar n}+g}{\w^2+E_\p^2} ,
\end{split}
\eeq
where $E_\p$ is the Bogoliubov excitation energy defined in Sec.~\ref{subsubsec_bog}. In the hydrodynamic regime $|\p|\ll p_c=\sqrt{2gm\bar n}$, 
\beq
\begin{split}
G_{nn}(p) &=  \frac{\bar n}{m} \frac{\p^2}{\w^2+c^2\p^2} , \\ 
G_{n\theta}(p) &= -\frac{\w}{\w^2+c^2\p^2} , \\
G_{\theta\theta}(p) &= \frac{mc^2}{\bar n}\frac{1}{\w^2+c^2\p^2} ,
\end{split}
\label{Gc}
\eeq
where $c=\sqrt{g\bar n/m}$ is the Bogoliubov sound mode velocity ($p_c=\sqrt{2}mc$).

\subsubsection{Exact hydrodynamic description} 
\label{subsubsec_hydro} 

In this section, we show that equations~(\ref{Gc}) are exact in the low-energy limit $|\p|,|\w|/c\ll p_c$ provided that $c$ is the exact sound mode velocity and $\bar n$ the actual mean density (which may differ from $\mu/g$). Let us consider the effective action $\Gamma[n,\theta]$ defined as the Legendre transform of the free energy $-\ln Z[J_n,J_\theta]$ ($J_n$ and $J_\theta$ are external sources linearly coupled to $n$ and $\theta$)~\cite{note14}. At zero temperature, $\Gamma[n,\theta]$ inherits Galilean invariance from the action (\ref{action6}). In a Galilean transformation (in imaginary time), $\r'=\r+i\v\tau$ and $\tau'=\tau$, the fields transform as 
\beq
\begin{split}
n'(x') &= n(x)  , \\
\theta'(x') &= \theta(x) - \frac{i}{2} m\v^2\tau - m\v\cdot\r , 
\end{split}
\eeq
where $x'=(\r',\tau')$. $n(x)$, $\nablabf n(x)$ and $i\dtau\theta+\frac{1}{2m}(\nablabf\theta)^2$ are Galilean invariant (but $\dtau n(x)$ is not). $\nablabf^2\theta$ is also invariant but is odd under time-reversal symmetry. Thus, to second order in derivatives, the most general effective action compatible with Galilean invariance and time-reversal symmetry reads 
\begin{align}
\Gamma[n,\theta] ={}& \int dx \biggl\lbrace \frac{Y(n)}{8m} (\nablabf n)^2 + U(n) \nonumber \\ & + \sum_{p=1}^2 c_p(n) \Bigl[ i\dtau\theta+\frac{1}{2m}(\nablabf\theta)^2 \Bigr]^p \biggr\rbrace , 
\end{align}
up to an additive (field-independent) term. $Y(n)$, $U(n)$ and $c_p(n)$ are arbitrary functions of $n$. 

To determine $c_p(n)$, we now consider the system in the presence of a fictitious vector potential $(A_0,\A)$, 
\begin{align}
S[n,\theta;A_\mu] = \int dx \biggl[ & in (\dtau\theta-A_0) + \frac{n}{2m}(\nablabf\theta-\A)^2 \nonumber \\ & + \frac{(\nablabf n)^2}{8mn} - \mu n + \frac{g}{2} n^2 \biggr] .
\end{align} 
The action is invariant under the local U(1) transformation $\theta\to\theta+\alpha$ and $A_\mu\to A_\mu+\partial_\mu \alpha$ where $\alpha(x)$ is an arbitrary phase. By requiring that $\Gamma[n,\theta;A_\mu]=\Gamma[n,\theta+\alpha;A_\mu+\partial_\mu \alpha]$ shares the same invariance, we deduce 
\begin{multline}
\Gamma[n,\theta;A_\mu] = \int dx \biggl\lbrace \frac{Y(n)}{8m} (\nablabf n)^2 + U(n) \\ 
+ \sum_{p=1}^2 c_p(n) \Bigl[ i\dtau\theta-iA_0+\frac{1}{2m}(\nablabf\theta-\A)^2 \Bigr]^p \biggr\rbrace . 
\end{multline}
Noting that 
\beq
n(x) = \frac{\delta \ln Z[J_n,J_\theta;A_\mu]}{i\delta A_0(x)} = - \frac{\delta \Gamma[n,\theta;A_\mu]}{i\delta A_0(x)}, 
\eeq
we must have $c_1(n)=n$ and $c_p(n)=0$ for $p\geq 2$. We conclude that 
\begin{align}
\Gamma[n,\theta] = \int dx \biggl\lbrace & \frac{Y(n)}{8m} (\nablabf n)^2 + U(n) \nonumber \\ & + n \left[ i\dtau\theta+\frac{(\nablabf\theta)^2}{2m} \right] \biggr\rbrace 
\label{gamma}
\end{align} 
to second order in derivatives. 

From (\ref{gamma}), we obtain the two-point vertex in constant fields $n(x)=\bar n$ and $\theta(x)=\const$ (with $\bar n$ the actual boson density), 
\begin{align}
\Gamma^{(2)}(p) &= \left( 
\begin{array}{cc} 
\Gamma^{(2)}_{nn}(p) & \Gamma^{(2)}_{n\theta}(p) \\
\Gamma^{(2)}_{\theta n}(p) & \Gamma^{(2)}_{\theta\theta}(p)
\end{array}
 \right) \nonumber \\ 
&= \left( 
\begin{array}{lr} 
 \frac{Y(\bar n)}{4m} \p^2 + U''(\bar n) & \w \\
-\w & \frac{\bar n}{m} \p^2
\end{array}
\right) . 
\end{align} 
By inverting $\Gamma^{(2)}(p)$, we recover the propagators (\ref{Gc}) in the small momentum limit $|\p|\ll p_c =[4mU''(\bar n)/Y(\bar n)]^{1/2}$ but with a sound mode velocity $c$ given by 
\beq
c = \sqrt{\frac{\bar n U''(\bar n)}{m}} .
\eeq
Noting that the compressibility $\kappa= \bar n^{-2}d\bar n/d\mu$ can also be expressed as~\cite{note13}
\beq
\kappa = \frac{1}{\bar n^2 U''(\bar n)} ,
\label{kappa}
\eeq
we conclude that the Bogoliubov sound mode velocity $c$ is equal to the macroscopic sound velocity $(m\bar n\kappa)^{-1/2}$. Moreover, since the superfluid density $n_s$ is defined by $\Gamma^{(2)}_{22}(\p,0)=\frac{n_s}{m}\p^2$ for $\p\to 0$~\cite{Dupuis09b}, we find that at zero temperature $n_s=\bar n$ is given by the fluid density~\cite{Gavoret64}.

\subsubsection{Normal and anomalous propagators}
\label{subsubsec_bosons_gnan}

To compute the propagator of the $\psi$ field, we write 
\beq
\psi(x) = \sqrt{n_0+\delta n(x)}e^{i\theta(x)}, 
\eeq
where $n_0=|\mean{\psi(x)}|^2=|\mean{\sqrt{n(x)}e^{i\theta(x)}}|^2$ is the condensate density. For a weakly interacting superfluid, $n_0\simeq \bar n$, and we expect the fluctuations $\delta n$ to be small. Let us assume that the superfluid order parameter $\mean{\psi(x)}=\sqrt{n_0}$ is real. Transverse and longitudinal fluctuations are then expressed as
\beq
\begin{split}
\delta\psi_2 &= \sqrt{2n_0}\theta + \cdots \\ 
\delta\psi_1 &= \frac{\delta n}{\sqrt{2n_0}} - \sqrt{\frac{n_0}{2}} \theta^2 + \cdots 
\end{split}
\eeq
where the ellipses stand for subleading contributions to the low-energy behavior of the correlation functions. For the transverse propagator, we obtain 
\beq
G_{22}(p) \simeq 2n_0 G_{\theta\theta}(p) = \frac{2n_0mc^2}{\bar n}\frac{1}{\w^2+c^2\p^2}
\label{G22}
\eeq
to leading order in the hydrodynamic regime, while
\beq
G_{12}(p) \simeq G_{n\theta}(p) = - \frac{\w}{\w^2+c^2\p^2} . 
\eeq
The longitudinal propagator is given by 
\begin{align}
G_{11}(x) &= \frac{1}{2n_0} G_{nn}(x) + \frac{n_0}{2} \mean{\theta(x)^2\theta(0)^2}_c \nonumber \\ 
&= \frac{1}{2n_0} G_{nn}(x) + n_0 G_{\theta\theta}(x)^2 , 
\end{align} 
where the second line is obtained using Wick's theorem (which is justified since the Goldstone (phase) mode is effectively non-interacting in the hydrodynamic limit). In Fourier space, 
\begin{align}
G_{11}(p) ={}& \frac{\bar n}{2mn_0}\frac{\p^2}{\w^2+c^2\p^2} + n_0 G_{\theta\theta}\star G_{\theta\theta}(p) ,
\label{G11}
\end{align} 
where
\beq 
G_{\theta\theta}\star G_{\theta\theta}(p) = \int_q G_{\theta\theta}(q)G_{\theta\theta}(p+q)
\eeq
with the dominant contribution to the integral coming from momenta $|\q|\lesssim p_c$ and frequencies $|\w'|/c\lesssim p_c$. Using (\ref{mint}), we find 
\begin{align}
\lefteqn{ G_{\theta\theta}\star G_{\theta\theta}(p)} \hspace{0.5cm}  & \nonumber \\ &= \llbrace 
\begin{array}{lcc}
 A_{d+1} c\left(\frac{m}{\bar n}\right)^2 \left(\p^2+\frac{\w^2}{c^2}\right)^{(d-3)/2} & \mbox{if} & d<3 , \\ 
\frac{A_4}{2} c \left(\frac{m}{\bar n}\right)^2 \ln \left(\frac{p_c^2}{\p^2+\frac{\w^2}{c^2}}\right)  & \mbox{if} & d=3 .
\end{array}
\right. 
\end{align}
By comparing the two terms in the rhs of (\ref{G11}) with $|\p|=p_G$ and $|\w|=cp_G$, we recover the Ginzburg scale (\ref{pG_est}). For $|\p|,|\w|/c\gg p_G$, the last term in the rhs of (\ref{G11}) can be neglected and we reproduce the result of the Bogoliubov theory (noting that $\bar n\simeq n_0$), while for $|\p|,|\w|/c\ll p_G$, $G_{11}(p)\sim 1/(\w^2+c^2\p^2)^{(3-d)/2}$ is dominated by phase fluctuations. The longitudinal susceptibility $G_{11}(\p,i\w=0)\sim 1/|\p|^{3-d}$ for $\p\to 0$ in contrast to the Bogoliubov approximation $G_{11}(\p,i\w=0)=1/2mc^2$. 

From these results, we deduce the hydrodynamic behavior of the normal propagator,
\begin{align}
G_{\rm n}(p) ={}& - \half \left[ G_{11}(p) -2iG_{12}(p) + G_{22}(p)\right] \nonumber \\ 
={}& - \frac{n_0mc^2}{\bar n} \frac{1}{\w^2+c^2\p^2} \nonumber \\ &
- \frac{i\w}{\w^2+c^2\p^2} - \half G_{11}(p) ,
\label{Gn}
\end{align}
as well as that of the anomalous propagator, 
\begin{align}
G_{\rm an}(p) ={}& - \half \left[ G_{11}(p) - G_{22}(p)\right] \nonumber \\  
&= \frac{n_0mc^2}{\bar n} \frac{1}{\w^2+c^2\p^2} - \half G_{11}(p) , 
\label{Gan}
\end{align}
where $G_{11}(p)$ is given by (\ref{G11}). The leading order terms in (\ref{Gn}) and (\ref{Gan}) agree with the results of Gavoret and Nozi\`eres~\cite{Gavoret64} and are exact (see next section). The contribution of the diverging longitudinal correlation function was first identified by
Nepomnyashchii and Nepomnyashchii~\cite{Nepomnyashchii78}, and later in Refs.~\cite{Popov79,Weichman88,Giorgini92,Castellani97,Pistolesi04}.

\subsubsection{Normal and anomalous self-energies}
\label{subsubsec_bosons_self}

To compute the self-energies $\Sign(p)$ and $\Sigan(p)$, we use the relations
\beq
\begin{split}
\Sign(p) &= G_0^{-1}(p) - \frac{\Gn(-p)}{\Gn(p)\Gn(-p)-\Gan(p)^2} , \\
\Sigan(p) &= \frac{\Gan(p)}{\Gn(p)\Gn(-p)-\Gan(p)^2} ,
\end{split}
\label{self}
 \eeq
with
\begin{align}
\lefteqn{ \Gn(p)\Gn(-p)-\Gan(p)^2} \hspace{0.75cm} & \nonumber \\ &= G_{11}(p)G_{22}(p)+G_{12}(p)^2  & \nonumber \\ 
&= G_{22}(p) \biggl[ n_0 G_{\theta\theta}\star G_{\theta\theta}(p) + \frac{\bar n}{2n_0mc^2} \biggr]. 
\end{align} 
Setting 
\beq
\begin{split}
\Gn(p) &\simeq - \half G_{22}(p) , \\
\Gan(p) &\simeq \half G_{22}(p) ,
\end{split}
\label{Gnan}
\eeq 
in the numerator of Eqs.~(\ref{self}), we obtain
\begin{align}
\Sigan(p) &= \Sign(p) - G_0^{-1}(p) \nonumber \\ &= \llbrace
\begin{array}{lcc} 
\frac{\bar n^2}{2A_{d+1}c^{4-d}n_0 m^2} (\w^2+c^2\p^2)^{(3-d)/2} & \mbox{if} & d<3 , \\ 
\frac{\bar n^2}{A_4 c n_0 m^2} \left[\ln \left(\frac{c^2 p_c^2}{\w^2+c^2\p^2}\right)\right]^{-1} & \mbox{if} & d=3 ,
\end{array}
\right. 
\label{sigma5}
\end{align} 
in the infrared limit $|\p|,|\w|/c\ll p_G$, where $G_0^{-1}(p)=i\w-\eps_\p+\mu$. Equations~(\ref{sigma5}) agree with the exact results (\ref{sigma6}) and show that $\Sign(p)$ and $\Sigan(p)$ are dominated by non-analytic terms for $p\to 0$. This non-analyticity reflects the singular behavior of the longitudinal correlation function 
\beq
G_{11}(p) \simeq \frac{1}{2\Sigan(p)} 
\eeq
in the low-energy limit. 

It should be noted that the singularity of the self-energies is crucial to reconcile the existence of a sound mode with a linear dispersion and the vanishing of the anomalous self-energy $\Sigan(p=0)$~\cite{Nepomnyashchii75}. In the low-energy limit,
\beq
\begin{split} 
\Sigan(p) &= \Delta\Sigma(p) + \Sigant(p) , \\ 
\Sign(p)-G_0^{-1}(p) &= \Delta\Sigma(p) + \Signt(p) ,
\end{split}
\eeq
where $\Delta\Sigma(p)$ denotes the singular part (\ref{sigma5}) while $\Signt(p)$ and $\Sigant(p)$ are regular contributions of order $\p^2,\w^2$. Using $\Delta\Sigma(p)\gg \Signt(p)-G_0^{-1}(p),\Sigant(p)$ for $p\to 0$, by inverting (\ref{propa}) we obtain 
\beq
\begin{split}
G_{\rm n}(p) &\simeq -\frac{1}{2[\Signt(p)-\Sigant(p)]} , \\ 
G_{\rm an}(p) &\simeq \frac{1}{2[\Signt(p)-\Sigant(p)]} . 
\end{split}
\label{Gnan1}
\eeq
Since both $\Signt(p)$ and $\Sigant(p)$ can be expanded to order $\p^2,\w^2$, we conclude that equations~(\ref{Gnan1}) predict the existence of a sound mode with linear dispersion. Of course, Eqs.~(\ref{Gnan1}) are nothing but our previous equations~(\ref{G22}) and (\ref{Gnan}).

In deriving the low-energy expression (\ref{sigma5}) of the self-energies, we have assumed that the hydrodynamic description holds up to the momentum scale $p_c$ and ignored the contribution of the non-hydrodynamic modes. In Popov's original approach~\cite{Popov79}, one introduces a momentum cutoff $p_0$ satisfying $p_G\ll p_0\ll p_c$. Since $p_0\gg p_G$, modes with momenta $|\p|\geq p_0$ can be taken into account within standard perturbation theory (see Sec.~\ref{subsec_bosons_pt}). On the other hand, low-momentum modes $|\p|\leq p_0\ll p_c$ are naturally treated in the hydrodynamic approach discussed in this section. The final results are independent of $p_0$. The only difference with our results (\ref{sigma5}) is that $p_c$ in the expression of $\Sigan(p)$ for $d=3$ is replaced by a smaller momentum scale~\cite{note10}.

\subsection{The non-perturbative RG} 
\label{subsec_bosons_nprg} 

The NPRG approach to zero-temperature interacting bosons has been discussed in detail in Refs.~\cite{Castellani97,Pistolesi04,Dupuis07,Dupuis09a,Dupuis09b,Wetterich08,Floerchinger08,Sinner09,Sinner10}. Our aim in this section is to briefly summarize the main results~\cite{note7} while emphasizing the common points with the classical O($N$) model studied in Sec.~\ref{subsec_phi4_nprg}.

To implement the NPRG, we add to the action an infrared regulator term
\beq
\Delta S_k[\psi^*,\psi] = \sum_{p} \psi^*(p) R_k(p) \psi(p) , 
\label{irreg}
\eeq
which suppresses fluctuations with momentum/frequency below a characteristic scale $k$ but leaves high momentum/frequency modes unaffected. The average effective action is defined as 
\begin{align}
\Gamma_k[\phi^*,\phi] ={}& - \ln Z_k[J^*,J] + \sum_{p} [J^*(p)\phi(p)+\cc] \nonumber \\ & - \Delta S_k[\phi^*,\phi] ,
\end{align}
where $\phi(x) = \mean{\psi(x)}_J$ is the superfluid order parameter. $J$ denotes a complex external source that couples linearly to the boson field.  $\Gamma_k$ satisfies the RG equation (\ref{rgeq}). As in Sec.~\ref{subsec_phi4_nprg}, we choose the cutoff function $R_k$ such that all fluctuations are suppressed for $k=\Lambda$ (so that $\Gamma_\Lambda[\phi^*,\phi]=S[\phi^*,\phi]$) and $R_{k=0}(p)=0$. In practice, we take~\cite{Dupuis09b}
\beq
R_k(p) = \frac{Z_{A,k}}{2m} \left(\p^2+\frac{\w^2}{c_0^2} \right) r\left( \frac{\p^2}{k^2}+\frac{\w^2}{k^2c_0^2} \right) , 
\label{regdef}
\eeq
where $r(Y)=(e^Y-1)^{-1}$. The $k$-dependent variable $Z_{A,k}$ is defined below. A natural choice for the velocity $c_0$ would be the actual ($k$-dependent) velocity of the Goldstone mode. In the weak coupling limit, however, the Goldstone mode velocity renormalizes only weakly and is well approximated by the $k$-independent value $c_0=\sqrt{g\bar n/m}$.

\subsubsection{Derivative expansion and infrared behavior} 
\label{subsubsec_bosons_de} 

The infrared regulator ensures that the vertices are regular functions of $p$ for $|\p|\ll k$ and $|\w|/c\ll k$ even when they become singular functions of $(\p,i\w)$ at $k=0$ ($c\equiv c_k\simeq c_{k=0}$ is the velocity of the Goldstone mode). In the low-energy limit $|\p|,|\w|/c\ll k$, we can therefore use a derivative expansion of the average effective action. We consider the ansatz 
\begin{align}
\Gamma_k[\phi^*,\phi] = \int dx\Bigl[ & \phi^*\Bigl(Z_{C,k}\dtau - V_{A,k}\partial^2_\tau - \frac{Z_{A,k}}{2m} \nablabf^2 \Bigr) \phi \nonumber \\ & + \frac{\lamb_k}{2} (n-n_{0,k})^2 \Bigr]
\label{effaction} 
\end{align} 
($n=|\phi|^2$), which is similar to the one used in the classical O($N$) model. $n_{0,k}$ denotes the condensate density in the equilibrium state. Note that we have introduced a second-order time derivative term. Although not present in the initial average effective action $\Gamma_\Lambda$, we shall see that this term plays a crucial role when $d\leq 3$~\cite{Wetterich08,Dupuis07}. As pointed out in Sec.~\ref{subsec_phi4_nprg}, the derivative expansion gives access only to the low-energy limit $|\p|,|\w|/c\ll k$ of the correlation functions. It is however possible to extract the $p$ dependence of the correlation functions by stopping the flow at $k\sim (\p^2+\w^2/c^2)^{1/2}$~\cite{Dupuis09b}.

In a broken symmetry state with order parameter $\phi_1=\sqrt{2n_0}$, $\phi_2=0$, the two-point vertex is given by 
\beq
\begin{split}
\Gamma^{(2)}_{k,11}(p) &= V_{A,k}\w^2+Z_{A,k}\eps_\p + 2\lamb_{k}n_{0,k}  , \\ 
\Gamma^{(2)}_{k,22}(p) &= V_{A,k}\w^2+Z_{A,k}\eps_\p , \\ 
\Gamma^{(2)}_{k,12}(p) &= Z_{C,k}\w .
\end{split}
\label{gamma2}
\eeq
Using (\ref{Gb}), we then find
\begin{align}
\Sigma_{k,{\rm n}}(p) ={}& G_0^{-1}(p) + \half \left[\Gamma^{(2)}_{k,11}(p)+\Gamma^{(2)}_{k,22}(p)\right] - i\Gamma^{(2)}_{k,12}(p) \nonumber \\
={}& \mu + V_{A,k}\w^2 + (1- Z_{C,k})i\w \nonumber \\ &- (1-Z_{A,k})\eps_\p + \lamb_k n_{0,k} 
\end{align}
and 
\beq
\Sigma_{k,{\rm an}}(p) = \half \left[\Gamma^{(2)}_{k,11}(p)-\Gamma^{(2)}_{k,22}(p)\right] = \lamb_{k}n_{0,k} . 
\eeq
At the initial stage of the flow, $Z_{A,\Lambda}=Z_{C,\Lambda}=1$, $V_{A,\Lambda}=0$, $\lamb_\Lambda=g$   and $n_{0,\Lambda}=\mu/g$, which reproduces the results of the Bogoliubov approximation. 

Since the anomalous self-energy $\Sigma_{k=0,{\rm an}}(p)\sim (\w^2+c^2\p^2)^{(3-d)/2}$ is singular for $|\p|,|\w|/c\ll p_G$ and $d\leq 3$, we expect $\Sigma_{k,{\rm an}}(p=0)\sim k^{3-d}$ for $k\ll p_G$ (given the equivalence between $k$ and $(\p^2+\w^2/c^2)^{1/2}$), i.e. 
\beq
\lamb_k \sim k^{3-d} . 
\label{lambk1}
\eeq
The hypothesis (\ref{lambk1}) is sufficient, when combined to Galilean and gauge invariances, to obtain the exact infrared behavior of the propagator. Furthermore we shall see that it is internally consistent. In the domain of validity of the derivative expansion, $|\p|^2,|\w|^2/c^2\ll k^2\ll k^{3-d}$ for $k\to 0$, one obtains from (\ref{gamma2})
\begin{equation}
\begin{split}
G_{k,11}(p) &= \frac{1}{2\lambda_kn_{0,k}} , \\ 
G_{k,22}(p) &= \frac{1}{V_{A,k}} \frac{1}{\w^2+c^2_k\p^2} , \\
G_{k,12}(p) &= -\frac{Z_{C,k}}{2\lambda_k n_{0,k} V_{A,k}} \frac{\w}{\w^2+c^2_k\p^2} ,
\end{split}
\label{de2}
\end{equation}
where
\begin{equation}
c_k = \left( \frac{Z_{A,k}/2m}{V_{A,k}+Z_{C,k}^2/2\lambda_k n_{0,k}} \right)^{1/2} 
\label{cdef} 
\end{equation}
is the velocity of the Goldstone mode. From (\ref{lambk1}) and (\ref{de2}), we recover the divergence of the longitudinal susceptibility if we identify $k$ with $(\p^2+\w^2/c^2)^{1/2}$.

The parameters $Z_{A,k}$, $Z_{C,k}$ and $V_{A,k}$ can be related to thermodynamic quantities using Ward identities~\cite{Gavoret64,Huang64,Pistolesi04,Dupuis09b},
\begin{equation}
\begin{split}
n_{s,k} &= Z_{A,k} n_{0,k} = \bar n_k , \\
V_{A,k} &= - \frac{1}{2n_{0,k}} \frac{\partial^2 U_k}{\partial\mu^2}\biggl|_{n_{0,k}} , \\ 
Z_{C,k} &= - \frac{\partial^2 U_k}{\partial n\partial\mu}\biggl|_{n_{0,k}} = \lambda_k \frac{dn_{0,k}}{d\mu} ,
\end{split}
\label{ward1} 
\end{equation}
where $\bar n_k$ is the mean boson density and $n_{s,k}$ the superfluid density. Here we consider the effective potential $U_k$ as a function of the two independent variables $n$ and $\mu$. The first of equations~(\ref{ward1}) states that in a Galilean invariant superfluid at zero temperature, the superfluid density is given by the full density of the fluid~\cite{Gavoret64}. Equations~(\ref{ward1}) also imply that the Goldstone mode velocity $c_k$ coincides with the macroscopic sound velocity~\cite{Gavoret64,Pistolesi04,Dupuis09b}, i.e.
\begin{equation}
\frac{d\bar n_k}{d\mu} = \frac{\bar n_k}{mc_k^2} .
\end{equation}
Since thermodynamic quantities, including the condensate ``compressibility''  $dn_{0,k}/d\mu$ should remain finite in the $k\to 0$ limit, we deduce from (\ref{ward1}) that $Z_{C,k} \sim \lambda_k \sim k^{3-d}$ vanishes in the infrared limit, and 
\begin{equation}
\lim_{k\to 0} c_k = \lim_{k\to 0} \left( \frac{Z_{A,k}}{2mV_{A,k}} \right)^{1/2} .
\label{velir}
\end{equation}
Both $Z_{A,k}=\bar n_k/n_{0,k}$ and the macroscopic sound velocity $c_k$ being finite at $k=0$, $V_{A,k}$ (which vanishes in the Bogoliubov approximation) takes a non-zero value when $k\to 0$. 

The suppression of $Z_{C,k}$, together with a finite value of $V_{A,k=0}$ shows that the effective action (\ref{effaction}) exhibits a ``relativistic'' invariance in the infrared limit and therefore becomes equivalent to that of the classical O(2) model in dimensions $d+1$~\cite{note4}. In the ordered phase, the coupling constant of this model vanishes as $\lamb_k\sim k^{4-(d+1)}$ (see Sec.~\ref{subsec_phi4_nprg}), which is nothing but our starting assumption (\ref{lambk1}). For $k\to 0$, the existence of a linear spectrum is due to the relativistic form of the average effective action (rather than a non-zero value of $\lambda_k n_{0,k}$ as in the Bogoliubov approximation). To neglect the term $Z_{C,k}\dtau$ in the average effective action (\ref{effaction}) (and therefore obtain a relativistic symmetry), it is necessary that $\lamb_k\gg k^2$~\cite{Dupuis09b}, a condition which is related to the singularity of the self-energies in the limit $p\to 0$. Thus we recover the fact that singular self-energies are crucial to obtain a linear spectrum in spite of the vanishing of the anomalous self-energy. 

To obtain the limit $k=0$ of the propagators (at fixed $p$), one should in principle stop the flow when $k\sim(\p^2+\w^2/c^2)^{1/2}$. Since thermodynamic quantities are not expected to flow in the infrared limit, they can be approximated by their $k=0$ values. As for the longitudinal correlation function, its value is obtained from the replacement $\lambda_k\to C(\w^2+c^2\p^2)^{(3-d)/2}$ (with $C$ a constant). From (\ref{de2}) and (\ref{ward1}), we then deduce the exact infrared behavior of the normal and anomalous propagators (at $k=0$),
\beq
\begin{split}
G_{\rm n}(p) ={}& - \frac{n_0mc^2}{\bar n} \frac{1}{\w^2+c^2\p^2} \\ &
- \frac{mc^2}{\bar n} \frac{dn_0}{d\mu} \frac{i\w}{\w^2+c^2\p^2} - \half G_{11}(p) , \\
G_{\rm an}(p) ={}& \frac{n_0mc^2}{\bar n} \frac{1}{\w^2+c^2\p^2} - \half G_{11}(p) , 
\end{split}
\label{de3}
\eeq
where
\begin{equation}
G_{11}(p) = \frac{1}{2n_0C (\w^2+c^2\p^2)^{(3-d)/2}} .
\label{de4}
\end{equation}
The hydrodynamic approach of Sec.~\ref{subsec_bosons_hydro} correctly predicts the leading terms of (\ref{de3}) but approximates $dn_0/d\mu$ by $\bar n/mc^2$. On the other hand, it gives an explicit expression of the coefficient $C$ in the longitudinal correlation function (\ref{de4}).

\subsubsection{RG flows}
\label{subsubsec_bosons_num} 

The conclusions of the preceding section can be obtained more rigorously from the RG equation satisfied by the average effective action. The dimensionless variables 
\beq
\begin{split} 
\tilde n_{0,k} &= k^{-d} Z_{C,k} n_{0,k} , \\
\tilde\lambda_k &= k^d\eps_k^{-1} Z_{A,k}^{-1}Z_{C,k}^{-1} \lambda_k , \\ 
\tilde V_{A,k} &= \eps_k Z_{A,k} Z_{C,k}^{-2} V_{A,k} ,
\end{split} 
\label{dimless_def}
\eeq
satisfy the RG equations 
\beq
\begin{split}
\dt \tilde n_{0,k} ={}& -(d+\eta_{C,k}) \tilde n_{0,k} + \frac{3}{2} \Itildell + \half \Itildett , \\
\dt \tilde \lambda_k ={}& (d-2+\eta_{A,k}+\eta_{C,k}) \tilde\lambda_k \\ & - \tilde\lambda_k^2 \bigl[9\Jtildellll(0) - 6\Jtildeltlt(0) + \Jtildetttt(0) \bigr] , \\ 
\eta_{A,k} ={}&  2 \tilde \lambda_k^2 \tilde n_{0,k}\frac{\partial}{\partial y} \bigl[\Jtildelltt(p)+\Jtildettll(p) \\ & +2\Jtildeltlt(p)\bigr]_{p=0} , \\ 
\eta_{C,k} ={}& -2\tilde\lambda_k^2\tilde n_{0,k} \frac{\partial}{\partial\tilde\w} \bigl[\Jtildettlt(p) - \Jtildelttt(p)  \\ & -3\Jtildelllt(p) + 3\Jtildeltll(p) \bigr]_{p=0} , \\ 
\dt \tilde V_{A,k} ={}& (2-\eta_{A,k}+2\eta_{C,k})\tilde V_{A,k} \\ & -2 \tilde\lambda_k^2\tilde n_{0,k} \frac{\partial}{\partial\tilde\w^2} \bigl[\Jtildelltt(p) + \Jtildettll(p) \\ & +2\Jtildeltlt(p)\bigr]_{p=0} ,
\end{split}
\label{flow_dim}
\eeq
where $\eta_{A,k}=-\dt\ln Z_{A,k}$, $\eta_{C,k}=-\dt\ln Z_{C,k}$, $y=\p^2/k^2$ and $\tilde\w=\w Z_{C,k}/Z_{A,k}\eps_k$. The definition of the threshold functions $\tilde I$ and $\tilde J$ can be found in Ref.~\cite{Dupuis09b}. 

\begin{figure}
\centerline{\includegraphics[width=6cm,clip]{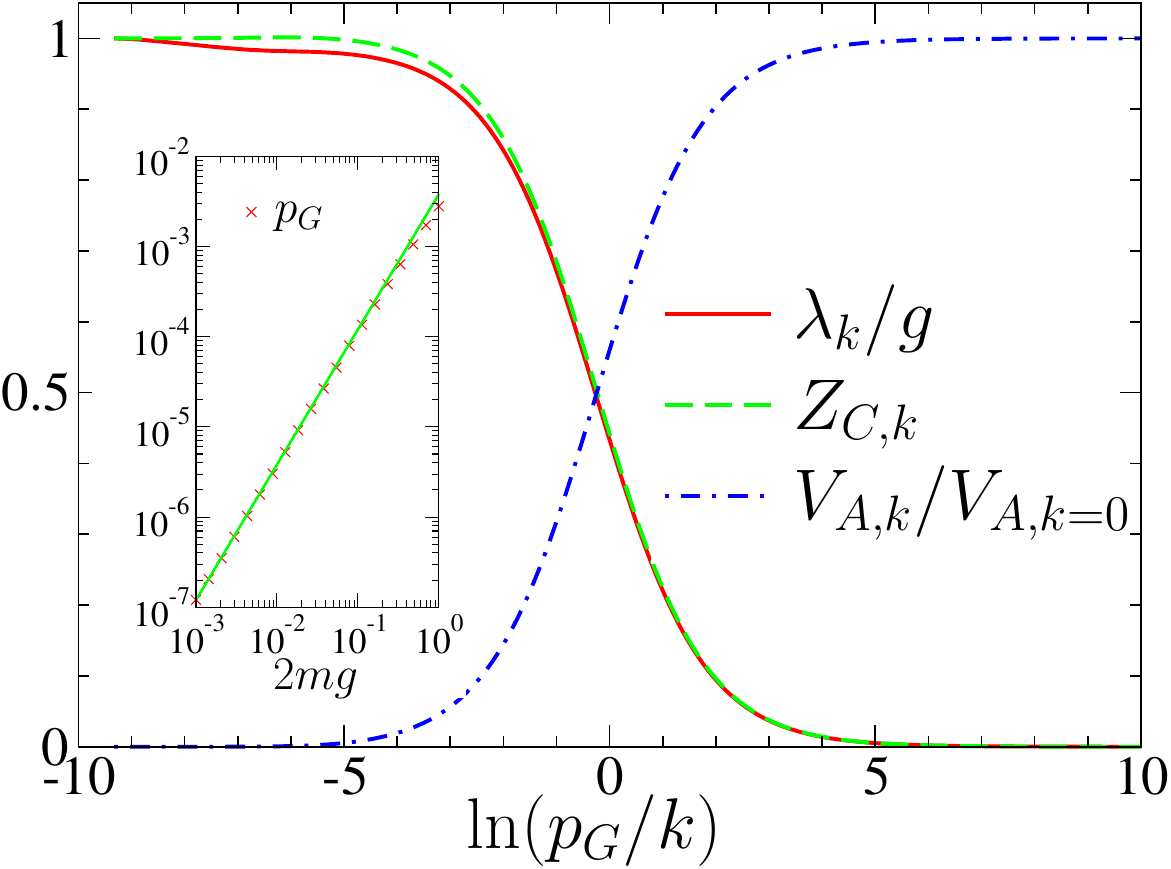}}
\caption{(Color online) $\lamb_k$, $Z_{C,k}$ and $V_{A,k}$ vs $\ln(p_G/k)$ where $p_G=\sqrt{(gm)^3\bar n}/4\pi$ for $\bar n=0.01$, $2mg=0.1$ and $d=2$ [$\ln(p_G/p_c)\simeq -5.87$]. The inset shows $p_G$ vs $2mg$ obtained from the criterion $V_{A,p_G}=V_{A,k=0}/2$ [the Green solid line is a fit to $p_G\sim (2mg)^{3/2}$].}
\label{fig_boson_flow_1} 
\centerline{\includegraphics[width=6cm,clip]{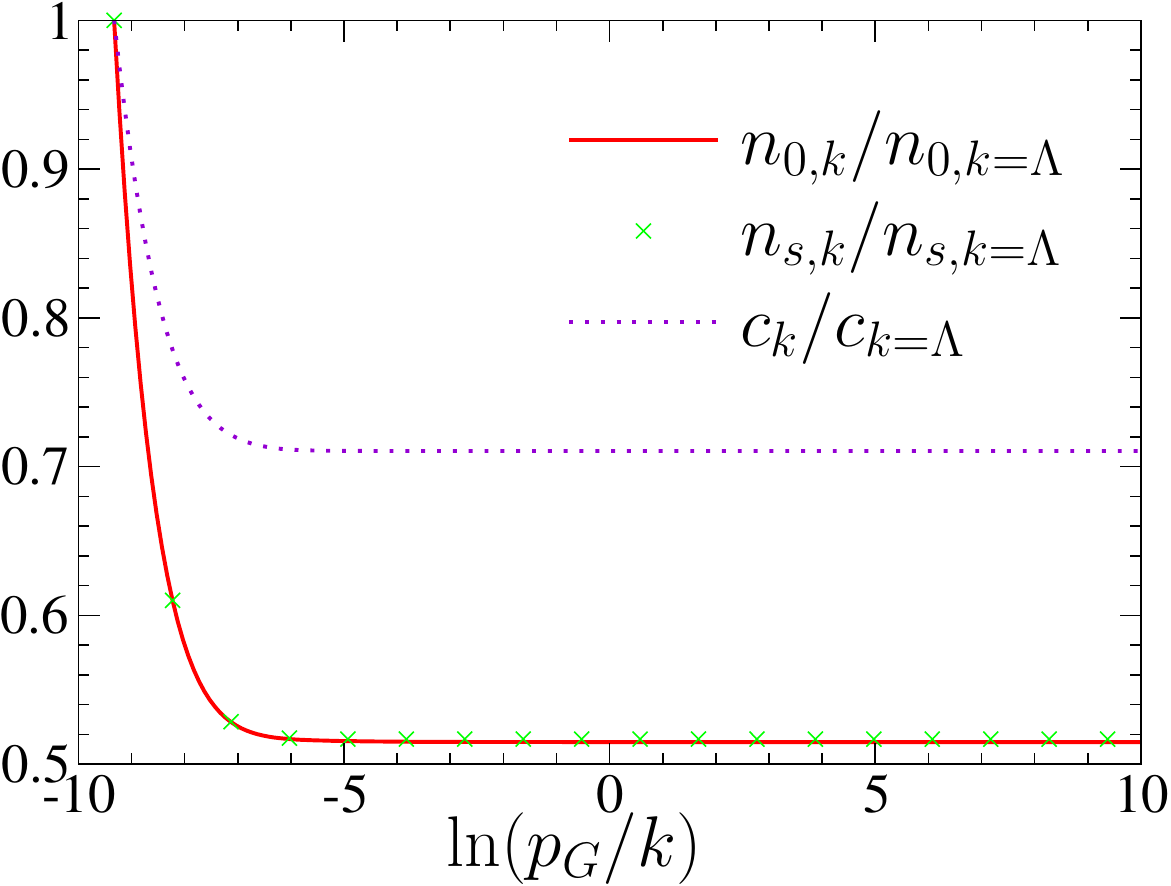}}
\caption{(Color online) Condensate density $n_{0,k}$, superfluid density $n_{s,k}$ and Goldstone mode velocity $c_k$ vs $\ln(p_G/k)$. The parameters are the same as in Fig.~\ref{fig_boson_flow_1}.} 
\label{fig_boson_flow_2}  
\centerline{\includegraphics[width=6cm,clip]{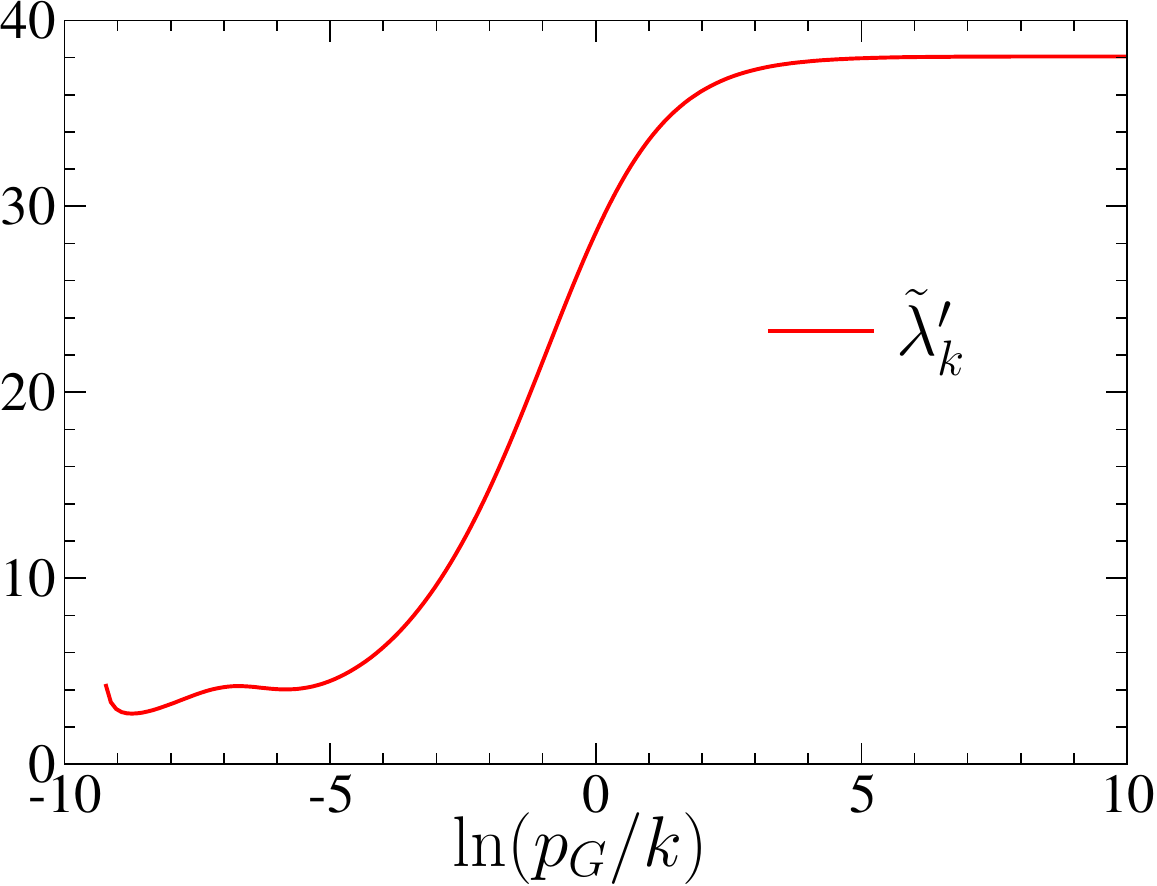}}
\caption{(Color online) $\tilde\lamb'_k$ vs $\ln(p_G/k)$ [Eq.~(\ref{lambp})]. The parameters are the same as in Fig.~\ref{fig_boson_flow_1}.}
\label{fig_boson_flow_3} 
\end{figure}

The flow of $\lamb_k$, $Z_{C,k}$ and $V_{A,k}$ is shown in Fig.~\ref{fig_boson_flow_1} for a two-dimensional system in the weak-coupling limit. We clearly see that the Bogoliubov approximation breaks down at a characteristic momentum scale $p_G\sim \sqrt{(gm)^3\bar n}$. In the Goldstone regime $k\ll p_G$, we find that both $\lamb_k$ and $Z_{C,k}$ vanish linearly with $k$ in agreement with the conclusions of Sec.~\ref{subsubsec_bosons_de}. Furthermore, $V_{A,k}$ takes a finite value in the limit $k\to 0$ in agreement with the limiting value (\ref{velir}) of the Goldstone mode velocity. Figure~\ref{fig_boson_flow_2} shows the behavior of the condensate density $n_{0,k}$, the superfluid density $n_{s,k}=Z_{A,k}n_{0,k}$ and the velocity $c_k$. Since $Z_{A,k=0}\simeq 1.004$, the mean boson density $\bar n_k=n_{s,k}$ is nearly equal to the condensate density $n_{0,k}$. Apart from a slight variation at the beginning of the flow, $n_{0,k}$, $n_{s,k}=Z_{A,k}n_{0,k}$ and $c_k$ do not change with $k$. In particular, they are not sensitive to the Ginzburg scale $p_G$. This result is quite remarkable for the Goldstone mode velocity $c_k$, whose expression (\ref{cdef}) involves the parameters $\lamb_k$, $Z_{C,k}$ and $V_{A,k}$, which all strongly vary when $k\sim p_G$. These findings are a nice illustration of the fact that the divergence of the longitudinal susceptibility does not affect local gauge invariant quantities~\cite{Pistolesi04,Dupuis09b}.

\subsubsection{Analytical results in the infrared limit} 
\label{subsubsec_bosons_analytic} 

In the Goldstone regime $k\ll p_G$, the physics is dominated by the Goldstone (phase) mode and longitudinal fluctuations can be ignored. If we take the regulator (\ref{regdef}) with $r(Y) = \frac{1-Y}{Y} \Theta(Y)$, the threshold functions $\tilde I$ and $\tilde J$ can be computed exactly and one obtains~\cite{Dupuis09b} 
\beq
\begin{split}
\dt\tilde n_{0,k} &= -(d+\eta_{C,k})\tilde n_{0,k} , \\ 
\dt\tilde\lambda_k &= (d-2+\eta_{C,k})\tilde\lambda_k + 8 \frac{v_{d+1}}{d+1} \frac{\tilde\lambda_k^2}{\tilde V_{A,k}^{1/2}} , \\ 
\eta_{C,k} &= - 8 \frac{v_{d+1}}{d+1} \frac{\tilde\lambda_k}{\tilde V_{A,k}^{1/2}} , \\ 
\dt\tilde V_{A,k} &=  (2+2\eta_{C,k}) \tilde V_{A,k} , 
\end{split} 
\label{eq_ir2}
\eeq
while $\eta_{A,k}\simeq 0$. The first and last of these equations can be rewritten as $n_{0,k}= n_{0,k=0}$ and $V_{A,k}=V_{A,k=0}$. From~(\ref{eq_ir2}), we deduce 
\beq
\begin{split}
\dt\tilde\lambda_k &= (1-\eps)\tilde\lambda_k , \\ 
\dt \eta_{C,k} &= -\eps\eta_{C,k} - \eta_{C,k}^2 , 
\end{split}
\eeq
where $\eps=3-d$. For $d<3$, this yields $\tilde\lamb_k\sim k^(1-\eps)$ and 
\beq
\lim_{k\to 0} \eta_{C,k} = -\eps , 
\eeq
i.e. $\lamb_k,Z_{C,k}\sim  k^\eps$ in agreement with the numerical results of Sec.~\ref{subsubsec_bosons_num} and the analysis of Sec.~\ref{subsubsec_bosons_de}. The anisotropy between time and space in the Goldstone regime $k\ll p_G$ (where the average effective action takes a relativistic form) can be eliminated by an appropriate rescaling of frequencies of fields. This leads to an isotropic relativistic model with dimensionless condensate density and coupling constant defined by~\cite{Dupuis09b}
\beq
\tilde n_{0,k}' = \sqrt{\tilde V_{A,k}} \tilde n_{0,k} , \qquad 
\tilde\lambda'_k = \frac{\tilde\lambda_k}{\sqrt{\tilde V_{A,k}}} .
\label{lambp} 
\eeq
$\tilde\lamb_k'$ satisfies the RG equation 
\beq
\dt\tilde\lamb_k' = -\eps \tilde\lamb'_k + 8 \frac{v_{d+1}}{d+1} \tilde\lamb'_k{}^2 , 
\label{rgeq3} 
\eeq
which is nothing but the RG equation of the coupling constant of the classical O(2) model in dimensions $d+1$ [Eq.~(\ref{rgeq2})]. The corresponding fixed point value can be deduced from (\ref{lambfp})~\cite{note5}. In the infrared limit, we find 
\beq
\lamb_k = k^{-d} (Z_{A,k}\eps_k)^{3/2} V_{A,k}^{1/2} \tilde\lamb_k' \sim k^\eps \tilde\lamb_k' 
\eeq
if we approximate $Z_{A,k}\simeq Z_{A,k=0}$ and $V_{A,k}\simeq V_{A,k=0}$. The vanishing of $\lamb_k\sim k^\eps$ and the divergence of the longitudinal susceptibility is therefore the consequence of the existence of a fixed point $\tilde\lamb'_k{}^*$ for the coupling constant of the effective ($d+1$)-dimensional O(2) model which describes the Goldstone regime $k\ll p_G$. To describe the entire hydrodynamic regime $k\ll p_c$, we should in principle relax the assumption $V_{A,k}\simeq V_{A,k=0}$, since $V_{A,k}$ strongly varies for $k\sim p_G$, which makes the analytical solution of the RG equations much more difficult. In Ref.~\cite{Sinner10}, it was shown that Eq.~(\ref{rgeq3}) is nevertheless in good agreement with the numerical solution of the flow equations in the entire hydrodynamic regime. We can then use (\ref{pG2}) to obtain the Ginzburg momentum scale
\beq
p_G \simeq  \left[\frac{8v_{d+1}g}{(d+1)\eps}\right]^{1/\eps}
\eeq
in the weak-coupling limit, which agrees with the results of Secs.~\ref{subsec_bosons_pt} and \ref{subsec_bosons_hydro}.

\section{Conclusion}

In conclusion, we have studied the classical linear O($N$) model and zero-temperature interacting bosons using a variety of techniques: perturbation theory, hydrodynamic approach, large-$N$ limit and NPRG. We have shown that in the weak-coupling limit these two systems can be described along similar lines. They are characterized by two momentum scales, the hydrodynamic scale (or healing scale for bosons) $p_c$ and the Ginzburg scale $p_G$. For momenta $|\p|\ll p_c$, we can use a hydrodynamic description in terms of amplitude and direction of the vector field $\varphibf$ in the O($N$) model, or density and phase in interacting boson systems. The hydrodynamic description allows us to derive the order parameter correlation function without encountering infrared divergences. In the Goldstone regime $|\p|\ll p_G$, amplitude (density) fluctuations play no role any more and both the transverse and longitudinal correlation functions are fully determined by direction (phase) fluctuations. In this momentum range, the coupling between transverse and longitudinal fluctuations leads to a divergence of the longitudinal susceptibility and singular self-energies. A direct computation of the order parameter correlation function (without relying on the hydrodynamic description) is possible, but one then has to solve the problem of infrared divergences which appear in perturbation theory when $|\p|\lesssim p_G$ and signal the breakdown of the Gaussian approximation. The NPRG provides a natural framework for such a calculation. In the case of bosons, it shows that in the Goldstone regime $|\p|,|\w|/c\ll p_G$, the system is described by an effective action with relativistic invariance similar to that of the $(d+1)$-dimensional classical O(2) model.

These strong similarities between the classical linear O($N$) model and zero-temperature interacting bosons disappear in the strong-coupling limit. For the O($N$) model, this limit corresponds to the critical regime near the phase transition, which has no direct analog in zero-temperature interacting boson systems. The only approach that one can hope to extend to strongly-correlated bosons is the NPRG. Recent progress in that direction, based on the Bose-Hubbard model, is reported in Ref.~\cite{note8}.

\begin{acknowledgments}
We would like to thank B. Svistunov for useful correspondence. 
\end{acknowledgments} 

\appendix

\section{Threshold functions} 
\label{sec_threshold}

The threshold functions appearing in the NPRG equations for the O($N$) model (Sec.~\ref{subsec_phi4_nprg}) are defined by 
\beq
\begin{split} 
I_\alpha &= - \int_\q \dot R(\q) G_{\alpha}^2(\q) , \\
J_{\alpha\beta}(\p)  &= - \int_\q \dot R(\q) G_{\alpha}^2(\q) G_{\beta}(\p+\q) , \\
J_{\alpha\beta}'(\p) &= \partial_{\p^2} J_{\alpha\beta}(\p) .
\end{split}
\eeq
where $\alpha,\beta\in \lbrace l,t\rbrace$. To alleviate the notations, we drop the $k$ index. In dimensionless form, 
\beq
\begin{split}
\tilde I_\alpha ={}& 2v_d \int_0^\infty dy\, y^{d/2}(\eta r+2yr') \tilde G_{\alpha}^2 , \\
\tilde J_{\alpha\beta}(0) ={}& 2v_d \int_0^\infty dy\, y^{d/2}(\eta r+2yr') \tilde G_{\alpha}^2 \tilde G_{\beta} , \\ 
\tilde J'_{\alpha\beta}(0) ={}& 4 \frac{v_d}{d} \int_0^\infty dy\, y^{d/2} \bigl\lbrace [\eta r+(\eta+4)yr'+2y^2r''] \tilde G_\alpha^2 \\ & - 2 (1+r+yr')(\eta r+2yr') \tilde G^3_\alpha \bigr\rbrace (1+r+yr') \tilde G_\beta^2 ,
\end{split}
\eeq
where 
\beq
\begin{split}
\tilde G_l &= \frac{1}{y(1+r)+2\tilde\lamb\tilde\rho_0} , \\
\tilde G_t &= \frac{1}{y(1+r)} , 
\end{split}
\eeq
and we have written the cutoff function as $R_k(\p)=Z_k\p^2 r(y)$ with $y=\p^2/k^2$ and $r(y)$ a $k$ independent function. For the theta cutoff function introduce in Sec.~\ref{subsubsec_phi4_rgeq}, $r=\frac{1-y}{y}\Theta(1-y)$, and the threshold functions can be computed analytically 
\beq
\begin{split}
\tilde I_{\alpha} &= -8\frac{v_d}{d} \left(1-\frac{\eta}{d+2}\right) \tilde A_\alpha^2 , \\ 
\tilde J_{\alpha\beta}(0) &= -8\frac{v_d}{d} \left(1-\frac{\eta}{d+2}\right) \tilde A_\alpha^2 \tilde A_\beta , \\ 
\tilde J'_{\alpha\beta}(0) &= 4 \frac{v_d}{d} \tilde A_l^2 ,
\end{split}
\eeq
where 
\beq
\tilde A_l = \frac{1}{1+2\tilde\lamb\tilde\rho_0} , \quad
\tilde A_t = 1 .
\eeq

%



\begin{thebibliography}{10}
\bibitem{note1}%
  \BibitemOpen
  \bibinfo {note} {Note that the measure
  $\calD[\sig]\delta(\sig^1+\pibf^2-1)=\calD[\sig] \frac{1}{2\sig}
  \delta(\sig-\sqrt{1-\pibf^2})$ does not contribute to the action
  (\ref{action2}) to leading order in the non-linear sigma model coupling
  constant $1/\rho_0^2$~\cite{Zinn_book_2}.}%
  \bibAnnoteFile{Stop}{note1}%
\bibitem{note9}%
  \BibitemOpen
  \bibinfo {note} {We use the same notations for the fields $\sig$ and $\lamb$
  and their classical values appearing in the effective action $\Gamma$.}%
  \bibAnnoteFile{Stop}{note9}%
\bibitem{note15} The vanishing of the anomalous dimension to leading order in $1/N$ is one of the main limitations of the large-$N$ approach.%
\bibitem{note3}%
  \BibitemOpen
  \bibinfo {note} {The result $\eta^*\simeq 0.054$ for the anomalous dimension
  is rather far from the best estimates $\eta^*\simeq 0.035$ for the O(3) model
  in three dimensions~\cite{Guida98} but can be improved by considering both
  $U_k(\rho)$ and $Z_k(\rho)$ as functions of $\rho$ with no further
  truncations: see Refs.~\cite{Tetradis94,Canet03}.}%
  \bibAnnoteFile{Stop}{note3}%
\bibitem{note14}%
  \BibitemOpen
  \bibinfo {note} {We use the same notations for the fields $n$ and $\theta$
  and their classical values appearing in the effective action $\Gamma$.}%
  \bibAnnoteFile{Stop}{note14}%
\bibitem{note13}%
  \BibitemOpen
  \bibinfo {note} {Equation~(\ref{kappa}) is the usual expression of the
  compressibility of a system with free energy density $F/V=U(\bar n)$ in the
  canonical ensemble.}%
  \bibAnnoteFile{Stop}{note13}%
\bibitem{note10}%
  \BibitemOpen
  \bibinfo {note} {For $d=2$, our resuls for $G_{11}(p)$ and $\Sigan(p)$ differ
  from Popov's~\cite{Popov79} by a factor of 2 but agree with those of
  Ref.~\cite{Giorgini92}.}%
  \bibAnnoteFile{Stop}{note10}%
\bibitem{note7}%
  \BibitemOpen
  \bibinfo {note} {We do not discuss the computation of the single-particle
  spectral function~\cite{Dupuis09a,Dupuis09b,Sinner09,Sinner10}.}%
  \bibAnnoteFile{Stop}{note7}%
\bibitem{note4}%
  \BibitemOpen
  \bibinfo {note} {To make this statement more rigorous, one must show that
  $Z_{C,k}$ disappears from the low-energy limit of the
  vertices~\cite{Dupuis09b}.}%
  \bibAnnoteFile{Stop}{note4}%
\bibitem{note5}%
  \BibitemOpen
  \bibinfo {note} {Note that the fixed point value $\tilde\lamb_k'{}^*$ deduced
  from Fig.~\ref{fig_boson_flow_3} differs from the result (\ref{lambfp}) since
  we have used different cutoff functions $R_k$ in the numerics and the
  analytical analysis of Sec.~\ref{subsubsec_bosons_analytic}.}%
  \bibAnnoteFile{Stop}{note5}%
\end{thebibliography}

\end{document}